\documentclass[12pt]{article}
\usepackage{dsfont}
\usepackage{graphicx}
\usepackage{amsmath}
\usepackage{amssymb}
\newtheorem{remark}{Remark}

\begin{document}
\title{Model reduction of cavity nonlinear optics for photonic logic: A quasi-principal components approach\thanks{Research supported by the Australian Research Council under Discovery Project grant DP130104191.}}
\author{Zhan Shi and Hendra I. Nurdin % <-this % stops a space
	%\thanks{This work was not supported by any organization}% 
	\thanks{
		Z. Shi and H. I. Nurdin are with School of Electrical Engineering and 
		Telecommunications,  UNSW Australia,  
		Sydney NSW 2052, Australia (e-mail: zhan.shi@sunsw.edu.au,  h.nurdin@unsw.edu.au).} 
}
\maketitle

\begin{abstract}
Kerr nonlinear cavities displaying optical thresholding have been proposed for the realization of ultra-low power photonic logic gates. In the ultra-low photon number regime, corresponding to energy levels in the attojoule scale, quantum input-output models become important to study the effect of unavoidable quantum fluctuations on the performance of such logic gates. However, being a quantum anharmonic oscillator, a Kerr-cavity  has an infinite dimensional Hilbert space spanned by the Fock states of the oscillator. This poses a challenge to simulate and analyze photonic logic gates and circuits composed of multiple Kerr nonlinearities. For simulation, the Hilbert of the oscillator is typically truncated to the span of only a finite number of Fock states. This paper develops a quasi-principal components approach to identify important subspaces of a Kerr-cavity Hilbert space and exploits it to construct an approximate reduced model of the Kerr-cavity on a smaller Hilbert space. Using this approach, we find a reduced dimension model with a Hilbert space dimension of 15 that can closely match the magnitudes of the mean transmitted and reflected output fields of a conventional truncated Fock state model of dimension 75, when driven by an input coherent field that switches between two levels. For the same input, the reduced model also closely matches the magnitudes of the mean output fields of  Kerr-cavity-based AND and NOT gates and a NAND latch obtained from simulation of the full 75 dimension model.  
%
%Valid PACS numbers may be entered using the \verb+\pacs{#1}+ command.
\end{abstract}

\maketitle

\section{\label{sec:intro} Introduction}
The last two decades have witnessed intense efforts into the engineering of quantum devices that are capable of quantum information processing, promising to revolutionize capabilities in computing, sensing, and communication with future quantum computers, quantum sensors, and quantum communication systems \cite{NC00}. However, a research line exploring quantum  devices for classical information processing with light \cite{Mab09,Mabuchi2011} has also emerged as an intermediate step towards full-fledged quantum information processing and which is important in its own right.  This has been driven by a revival of interest in photonic computing using photonic logic gates, spurred on by experimental advances in nanophotonics. A major challenge facing current semiconductor chips as feature sizes of components continually shrink is that smaller and smaller metallic interconnects dissipate more and more heat, thus making the resulting devices increasingly energy inefficient. This has prompted the radical solution of replacing on-chip metallic interconnects with photonic buses, thus  replacing electrons as the traditional carrier of information in semiconductors with photons \cite{BKSWW07}. A natural extension of this line of thought is to make the entire chip all-optical, thereby replacing  all components on a chip, including transistors that implement logic operations, with photonic counterparts.

An exciting on-going research theme is the engineering of photonic devices as ultra-low power photonic logic gates. That is,  photonic logic gates that can operate in the regime of just a few photons (energy in the attojoule scale), see, e.g., \cite{Mab09,Mabuchi2011,Tezak2012,SPBTHM14,TKM16} and the references cited therein. However, at such low energy levels, the presence of quantum fluctuations/noise can seriously impair the performance of such devices, by causing spontaneous unwanted  switching of logical values. Therefore, despite the devices only being used for classical information processing, for low photon number applications it is important to use fully quantum mechanical models in their simulation and analysis  in order to capture the effects of quantum fluctuations. Photonic devices with traveling optical fields as inputs and outputs are accurately modeled using quantum stochastic differential equations (QSDEs), developed independently by Hudson and Parthasarathy \cite{HP84,KRP92,Mey95} and Gardiner and Collett \cite{GC85,GZ04}.  Individual photonic devices can be interconnected together to form cascaded systems \cite{Gough2009,Gard93,Carm93} and more complex  feedback interconnected systems \cite{Gough2009math,Gard93,Carm93}, which can again be modelled as a QSDE in the regime where the propagation time for the optical fields interconnecting the network components can be neglected with respect to the dynamical time scales of the network components. A freely available schematic capture  workflow software package called QHDL (Quantum Hardware Description Language)  has been developed to automate the calculation of parameters of the QSDE describing a quantum network, see \cite{Tezak2012} for an introduction. 
 
Nonlinear optical cavities with a Kerr nonlinear medium have been proposed as candidates for ultra-low power photonic logic gates. Such cavities are under the Kerr effect, that is, the refractive index of the Kerr nonlinear medium interacting with an incident light is dependent on the light intensity \cite{bBoyd2008}. For a certain range of the incident light intensity, the Kerr effect can lead to the presence of thresholding and bistability. Thresholding behaviour is manifested by a rapid increase in the output field intensity as the input field intensity is increased beyond a certain value, while bistability is an optical effect that a single value of the incident light intensity corresponds to two different intensity values of the cavity outgoing field \cite{Smith1979,Yanik2003}. 
By exploiting the optical thresholding, a single Kerr nonlinear optical cavity can be configured using all-optical feedback to form AND, NOT, and pseudo-NAND\footnote{Pseudo here is in the sense that this NAND gate require at least one of its two inputs to be at the HIGH logical value at any time.} gates, while two pseudo-NAND gates can be connected in a feedback loop to form a NAND latch \cite{Mabuchi2011,Tezak2012}. As alluded to above, these gates can be described using the QSDE formalism, and the QSDE derived by the network calculations shown in \cite{Tezak2012} (see also Appendix to this paper), which can be automated using QHDL. To evaluate their performance, simulation of the gates can be carried out employing the techniques of quantum trajectory theory \cite{Carm93}. However,  Kerr nonlinear cavities are anharmonic oscillators with an infinite-dimensional Hilbert space spanned by the Fock states $\{ |0\rangle, |1 \rangle, |2 \rangle,\ldots \}$. Thus, for simulation on a digital computer, this Hilbert space has to be necessarily truncated to a finite-dimensional one, $\mathcal{H}^{(N)}=\{ |0\rangle, |1 \rangle, |2 \rangle,\ldots,|N\rangle\}$ for some finite $N$ chosen to be sufficiently large to match the amplitude of the driving field, which we refer to as the ``full model''. In the work \cite{Tezak2012}, the simulation was carried out with $N=75$. When two Kerr-cavities are interconnected to form the NAND latch, the dimension of the Hilbert space becomes $N^2$. As several Kerr-cavity gates are interconnected to form circuits performing more complex logical operations, it is of interest to be able to simulate such circuits to assess the impact of quantum fluctuations on their performance. However, due to the exponential increase in the size of Hilbert space with the number of components, it is important to develop methods to reduce the dimension of the model and to efficiently simulate the circuit. The process of simplifying a dynamical model by reducing its dimension or number of variables is referred to as {\em model reduction}. For the Kerr NAND latch, an empirical model reduction method for the latch driven by an input switching between two values was proposed in \cite{Tezak2012}. The approach was based on identifying a classical continuous-time Markov chain that could be embedded in the dynamics of an input-output quantum system with a small finite-dimensional Hilbert space (chosen in \cite{Tezak2012} to be 38 for the NAND latch). The parameters of this Markov chain were empirically estimated by running a large number of quantum trajectory simulations of the full latch model of dimension $N^2=75^2$. However, such a reduced model required coupling to artificial additional vacuum noise channels that were not present in the original full model to facilitate switching between states of the classical Markov chain. In \cite{SPBTHM14}, a semiclassical approximation  was employed for efficient numerical simulation of circuits composed of multiple Kerr nonlinear cavities. The idea there is to approximate the master equation for the internal evolution of the circuit with a Fokker-Planck equation corresponding to a classical stochastic system that approximates the dynamics of the original quantum circuit. Although allowing to simulate a network with a significant number of Kerr components, such an approximation was found to be effective for higher photon numbers but become noticeably less accurate when the system is operating at lower photon numbers (``high'' and ``low'' here  are relative to the amplitude of the coherent field driving the circuit).  

In this paper, we develop a model reduction method for a single Kerr nonlinear cavity driven by a coherent field switching between two values representing two logic levels of HIGH and LOW. We exploit the idea of identifying quasi-principal components (vectors) of the Hilbert space of the full model. The quasi-principal vectors are extracted from the steady-state density operator of a single Kerr-cavity when driven by a vacuum field and another fixed coherent driving field chosen in the vicinity of the cavity's threshold region. The Hilbert space of the reduced model is then chosen as the span of a small subset of these vectors, selected according to an approximate simultaneous diagonalization criterion. The reduced model is not an embedding of a classical system in a quantum system nor is it  a semiclassical model, but is a genuine quantum model that is extracted using intrinsic features of the high order 75 dimensional model.   Moreover,  the reduced model does not require coupling to additional fictitious fields  that were not part of the full model. With this approach we construct a reduced model  for a single Kerr-cavity living on a finite-dimensional Hilbert space of dimension 15. This reduced model  is subsequently used to replace the full model in circuits realizing AND and NOT gates, and the NAND latch. It is demonstrated by numerical examples that reduced model is able to closely approximate the magnitude of the mean output field  dynamics of the gates and latch.   

The structure of this paper is as follows.  Section \ref{sec:prelim} gives a brief technical background of input-output open  quantum systems and their descriptions using QSDEs. In the two sections that then follow,  inspired by principal component analysis for statistical data  analysis and pattern recognition \cite{bJolliffe2002,Fuku90} and signal processing \cite{Clarke85,Jain89}, we  first introduce a new method based on a notion of quasi-principal components for obtaining a low-dimensional reduced quantum model of a Kerr nonlinear cavity, when the driving input field switches between two  values.  After that Section~\ref{sec:gates} constructs reduced quantum models  of photonic logic gates containing one or two Kerr-cavities by substituting the full cavity model with its reduced model. Finally, Section \ref{sec:conclu} summarizes the contributions of the paper and discusses some directions for future research. Details of the calculations associated with the models and quantum networks that are considered in the main text have been collected together in Appendix.

\section{Technical  background}
\label{sec:prelim}
\noindent \textbf{Notation.} This paper employs the following notations: $\imath$ denotes $\sqrt{-1}$. 
$\cdot^*$ denotes (i) the complex conjugate of a number, (ii) the conjugate transpose of a matrix, as well as (iii) the adjoint of an operator. 
$O_{m\times n}$ denotes an $m$ by $n$ zero matrix and $I_n$ denotes an $n$ by $n$ identity matrix, where we simply write $O_m$ if $m=n$, and drop the subscripts if the dimensions of the matrices can be inferred from context. 
$\rm{Im}[X] = (X-X^*)/(2\imath)$  for any complex number or operator $X$. 
The trace operator is  denoted by  $\rm{Tr[\cdot]}$ and tensor product is denoted by $\otimes$.  
$\delta_{ij}$ denotes the Kronecker delta and $\delta(t)$ denotes the Dirac delta function. 
Commutator $[a,b]$ and anti-commutator $\{a,b\}$ of two operators $a$ and $b$ act as $[a,b]=ab-ba$ and $\{a,b\}=ab+ba$.

We are concerned with a class of quantum networks consisting of input-output open quantum Markov subsystems as nodes, which are interconnected by traveling boson fields. The boson fields act as inputs to the systems and are scattered as output fields after their interaction with the system. An input-output open Markov quantum subsystem $G$ with $n$ incoming fields and $n$ outgoing fields is characterized by a triple $G=(S, L, H)$, referred as the set of system parameters or `SLH', where
\begin{eqnarray}
S&=&\left[\begin{array}{ccc}
S_{11} & \cdots & S_{1n} \\
\vdots & \ddots & \vdots \\
S_{n1} & \cdots & S_{nn}
\end{array}\right], ~~~
L=\left[\begin{array}{ccc}
L_{1}\\
\vdots\\
L_n
\end{array}\right]
\end{eqnarray}
\normalsize
are a unitary scattering matrix with operator entries and a vector of system-field coupling operators, respectively; $H$ is a Hermitian operator representing the Hamiltonian of the subsystem (without the fields) \cite{Gough2009,Gough2009math}. The operators $S_{jk}$, $L_j$ ($1 \leq j,k \leq n$) and $H$ 
%involved in the triple 
are defined on an infinite-dimensional Hilbert space $\mathcal{H}$, the Hilbert space for the node, referred as the initial space. In our case, the system of interest will be a quantum oscillator and the initial space will be the space ${\rm span}\{|0\rangle, |1 \rangle, \ldots, ..., |n \rangle,... \}$ spanned by the Fock states of the oscillator. The incoming fields are boson white noises described by field annihilation operators $\xi_t^{j}$ and field creation operators ${\xi_t^{j}}^*$ ($1 \leq j \leq n$), $j$ is an index to denote a particular field. The field operators satisfy the commutation relations
$\left[\xi_t^i, {\xi_s^j}^*\right]=\delta_{ij}\delta(t-s)$ and $\left[\xi_t^i, \xi_s^j\right]=0$ under the Markov approximation.
Annihilation, creation and gauge processes are defined as $A_t^j=\int_{0}^{t} \xi_s^j ds$, ${A_t^j}^*=\int_{0}^{t} {\xi_s^j }^*ds$ and $\Lambda_t^{jk}=\int_{0}^{t}{\xi_s^j }^*\xi_s^k ds$, respectively, which are adapted processes satisfying the quantum It\={o} rules $dA_t^jd{A_t^k}^*=\delta_{jk}dt$, $dA_t^jd\Lambda_t^{kl}=\delta_{jk}dA_t^l$, $d\Lambda_t^{ij}d{A_t^k}^*={\delta_{jk}dA_t^i}^*$ and $d\Lambda_t^{ij}d\Lambda_t^{kl}=\delta_{jk}d\Lambda_t^{il}$, where $dA_t^k=A_{t+dt}^k-A_t^k$, $d{A_t^k}^* ={A_{t+dt}^k}^*-{A_t^k}^*$ and $d \Lambda_{t}^{kl}=\Lambda_{t+dt}^{kl} -\Lambda_t^{kl} $.
The dynamics of the subsystem is given by the right Hudson-Parthasarahty quantum stochastic differential equation (QSDE)
\begin{eqnarray}
dU_{t} &=& \left(\sum_{j,k=1}^{n} \left(S_{jk}-\delta_{jk}\right) d\Lambda_{t}^{jk} + dA_t^* L - L^*SdA_t   -\left(\imath H + \frac{1}{2}L^*L \right)dt\right)U_t; ~U_0=I, 　
\end{eqnarray}
\normalsize
where $U_t$ is a unitary adapted process. After interaction with the system, the input annihilation processes $A_t=(A^1_t,A^2_t,\ldots,A^m_t)^{\top}$ are transformed to the output annihilation processes given by $Y_t = U_t^* A_t U_t$ and having the differential form $dY_t = L_t dt + S_t dA_t$, where $L_t =U_t^* L U_t$ and $S_t = U_t^*SU_t$. The output process $Y_t$ has two quadratures, $Y_t^Q = Y_t + Y_t^*$, called the amplitude quadrature, and $Y_t^P = -\imath Y_t + \imath Y_t^*$, called the phase quadrature. Each of these quadratures can be separately measured in the lab (but not simultaneously as they do not commute) by a process called homodyne detection.

In this paper, we first approximate  a given subsystem $G$, that will represent a Kerr nonlinear cavity and described in detail in the next section,  with a reduced model $G^{(r)}=(S^{(r)}, L^{(r)}, H^{(r)})$ that is defined on a finite-dimensional subspace $\mathcal{H}^{(r)} \subset \mathcal{H}$. The dynamics of the reduced subsystem is governed by a unitary adapted process $U^{(r)}_t$ satisfying the right Hudson-Parthasarathy  QSDE
\small
\begin{eqnarray}
dU_{t}^{(r)} &=& \left(\sum_{j,k=1}^{n} \left(S_{jk}^{(r)}-\delta_{jk}\right) d\Lambda_{t}^{jk}  - {L^{(r)}}^*S^{(r)}dA_t  + dA_t^* L^{(r)} -\left(\imath H^{(r)} + \frac{1}{2}{L^{(r)}}^*L^{(r)} \right)dt\right)U_t^{(r)}; \nonumber\\
U_0^{(r)}&=&I.　
\end{eqnarray}
\normalsize
The output annihilation process $Y_t^{(r)}$ of the reduced model is given by $Y_t^{(r)} = (U_t^{(r)})^* A_t U_t^{(r)}$ and satisfies $dY_t^{(t)} = L_t^{(r)}dt + S_t^{(r)} dA_t$, with $L_t^{(r)} = (U_t^{(r)})^* L U_t^{(r)}$ and $S_t^{(r)} =(U_t^{(r)})^* S U_t^{(t)}$.  

Once $G^{(r)}$ is obtained, we are interested in constructing a reduced model for the quantum network by replacing the original interconnected subsystem $G$ with $G^{(r)}$. In particular, quantum networks considered in this paper are a set of cavity nonlinear optical models of AND and NOT gates and a NAND latch which were proposed in \cite{Mabuchi2011}. These composite quantum models consist of at least one Kerr nonlinear optical ring cavity whose system operators are defined on the infinite-dimensional Hilbert space $\mathcal{H}$, and static linear optical devices such as beamsplitters and phase shifters.

%%%%%%%%%%%%%%%%%%%%%%%%%%%%%%%%%%%%%%%%%%%%%%%%%%%%%%%%%%%%%%%%%%%%%%%%%%%%%%%%%%%%%%%%%%%%%%%%%%%%%%%%%%%%%%%%%%%%%%%%%%%%%%%%%%%%%%%%%%%%%%%%%%%%%%%%%
%%%%%%%%%%%%%%%%%%%%%%%%%%%%%%%%%%%%%%%%%%%%%%%%%%%%%%%%%%%%%%%%%%%%%%%%%%%%%%%%%%%%%%%%%%%%%%%%%%%%%%%%%%%%%%%%%%%%%%%%%%%%%%%%%%%%%%%%%%%%%%%%%%%%%%%%%
\section{\label{sec:kerr_cavity} Model reduction of a Kerr-cavity}
\begin{figure}[htbp]
	\begin{center}
	\includegraphics[scale=0.5]{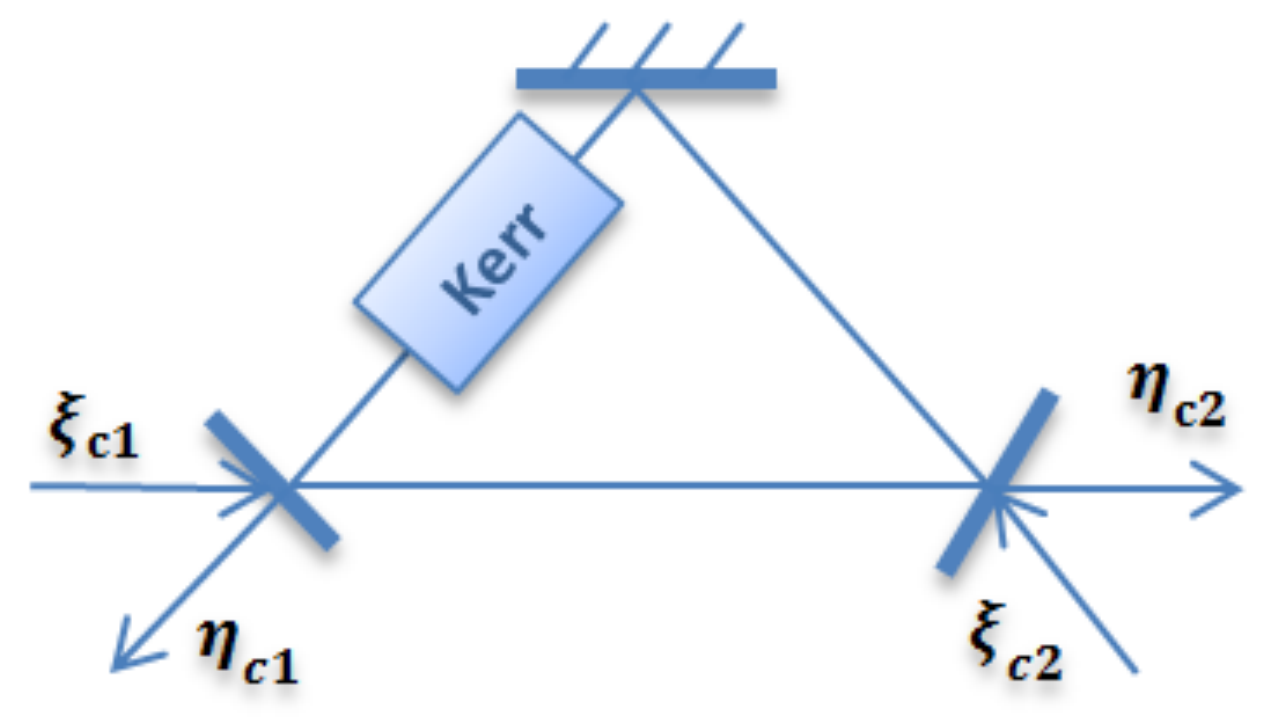}% Here is how to import EPS art
	\caption{\label{fig:Kerr} A Kerr-nonlinear cavity.}
	\end{center}
\end{figure}

\begin{figure}[htbp]
	\begin{center}
		\includegraphics[scale=0.45]{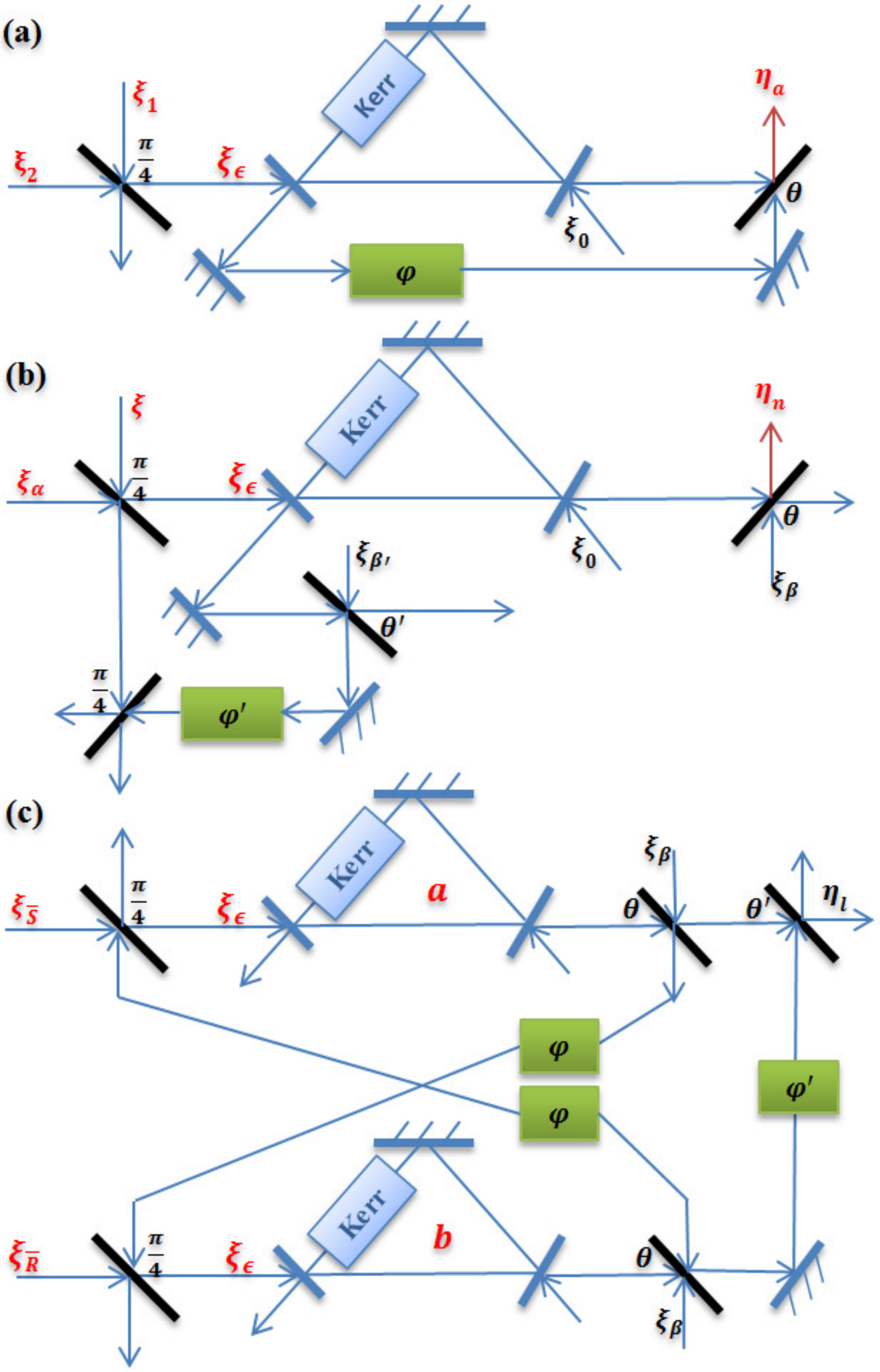}% Here is how to import EPS art
		\caption{\label{fig:gates} (a) An AND gate. (b) A NOT gate. (c) A NAND gate. The black lines are beam splitters and the green blocks denote phase shifters.}
	\end{center}
\end{figure}
A Kerr nonlinear optical ring cavity shown in Figure~\ref{fig:Kerr} is an open nonlinear input-output quantum system containing a Kerr medium with the nonlinear coefficient $\chi$ and an internal mode $a$. Its dynamics is governed by the system Hamiltonian $H_0^a=\Delta a^*a+ \chi a^*a^*aa$, where $\Delta$ is the detuning of the cavity resonance frequency with respect to a reference carrier frequency $\omega_r$\footnote{We work in a rotating frame at this carrier frequency.}.  The cavity has a fully reflecting mirror and two partially transmitting mirrors with damping rate $\kappa$. Two incoming fields $\xi_{c1}$ and $\xi_{c2}$ are coupled with the cavity mode $a$ via coupling operators $L_1=\sqrt{\kappa}a$ and $L_2=\sqrt{\kappa}a$, respectively. We assume $\xi_{c1}$ is a coherent field with amplitude $\epsilon$ referred to as the cavity driving field\footnote{This is in the rotating frame at frequency $\omega_r$. In the lab frame the amplitude is actually $\epsilon e^{-\imath \omega_r t}$.}, and $\xi_{c2}$ is in the vacuum state. 
The quantities of interest are the expectation values of the reflected outgoing field $\eta_{c1}$ and the transmitted outgoing field $\eta_{c2}$, that is,  $\left\langle \eta_{c1} \right\rangle = {\rm Tr }[\rho_t \sqrt{\kappa} a ]+\epsilon$ and  $\left\langle \eta_{c2} \right\rangle = {\rm Tr }[\rho_t \sqrt{\kappa} a ]$, where $\rho_t$ is the cavity density operator at time $t$, which can be calculated by the master equation
$
\frac{d}{dt}\rho_t =-\imath \left[H, \rho_{t} \right]+ 2\left(L \rho_t L^* - \frac{1}{2} L^*L\rho_{t} 　- \frac{1}{2} \rho_{t} L^*L\right), \nonumber
$
where $H=H_0^a+ \imath\sqrt{\kappa}\epsilon(a-a^*)$.

\begin{figure}[htbp]
	\begin{center}
	\includegraphics[scale=0.43]{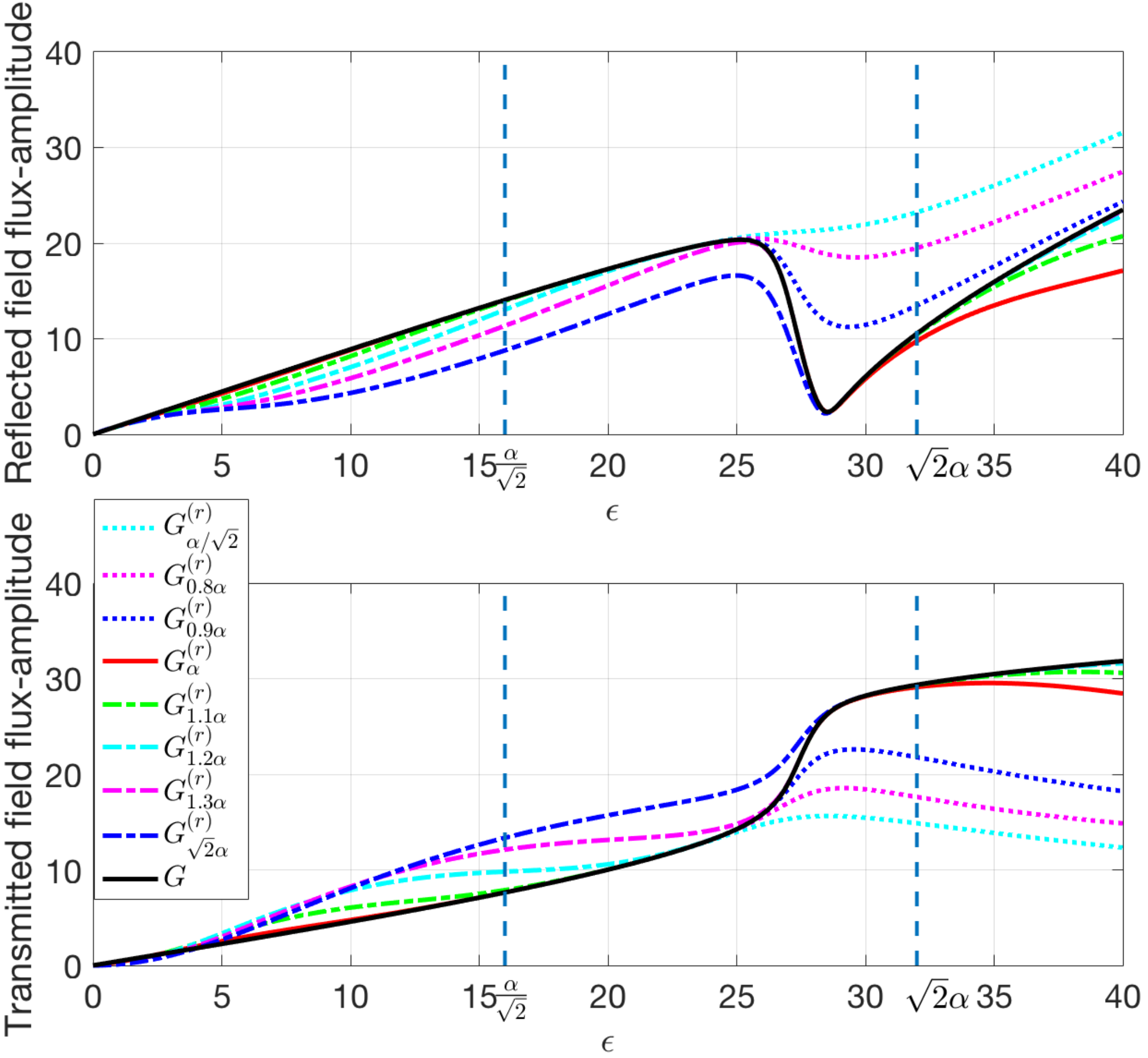}% Here is how to import EPS art
	\caption{\label{fig:cav_outputs} Comparison of steady-state reflected (top) and transmitted (bottom) magnitudes of the mean output fields of the full cavity model $G$ and JADE-based reduced models $G_\lambda^{(r)}$, with $\lambda \in \{\alpha/\sqrt{2}, 0.8 \alpha, 0.9\alpha, \alpha, 1.1\alpha,1.2\alpha,1.3\alpha,\sqrt{2} \alpha \}$ and $d=15$, as functions of $\epsilon$.}
	\end{center}
\end{figure}

Taking the same numerical parameters and units as in \cite{Mabuchi2011}, we set $\kappa=25$, $\Delta=50$ and $\chi=-\Delta/60$, and employ the Quantum Optics Toolbox for MATLAB (qotoolbox) to numerically compute the steady-state mean cavity transmitted and reflected outputs. We note that  these parameter values were specifically chosen in \cite{Mabuchi2011} to obtain  switching dynamics corresponding to stored energy in the range of tens of photons per cavity, as  this appears to represent the minimum operational energy scale that can be considered before quantum shot-noise fluctuations start to dominate the device physics. In numerical calculations, cavity operators are restricted to a truncated Hilbert space $\mathcal{H} ^{(N)}$, spanned by the Fock state basis  $\{\arrowvert n \rangle\}_{0\leq n \leq N-1}$ for some nonnegative integer $N$.  Following \cite{Tezak2012},  we set $N=75$ corresponding to a maximum intra-cavity photon number of 74. From now on, as specified in the introduction, the cavity model in $\mathcal{H}^{(N)}$ with $N=75$ will be referred to as the full model. Figure~\ref{fig:cav_outputs} shows the magnitudes of the mean flux-amplitudes of transmitted and reflected cavity outgoing fields (black solid line), with the cavity driving amplitude $\epsilon$ varying from $0$ to $40$.  As in \cite{Mabuchi2011}, the rapid change in output field intensity at around $\epsilon \approx 26$ is the thresholding behavior due to the Kerr effect \cite{Yanik2003}. 

\begin{figure}[htbp]
	\begin{center}
		\includegraphics[scale=0.4]{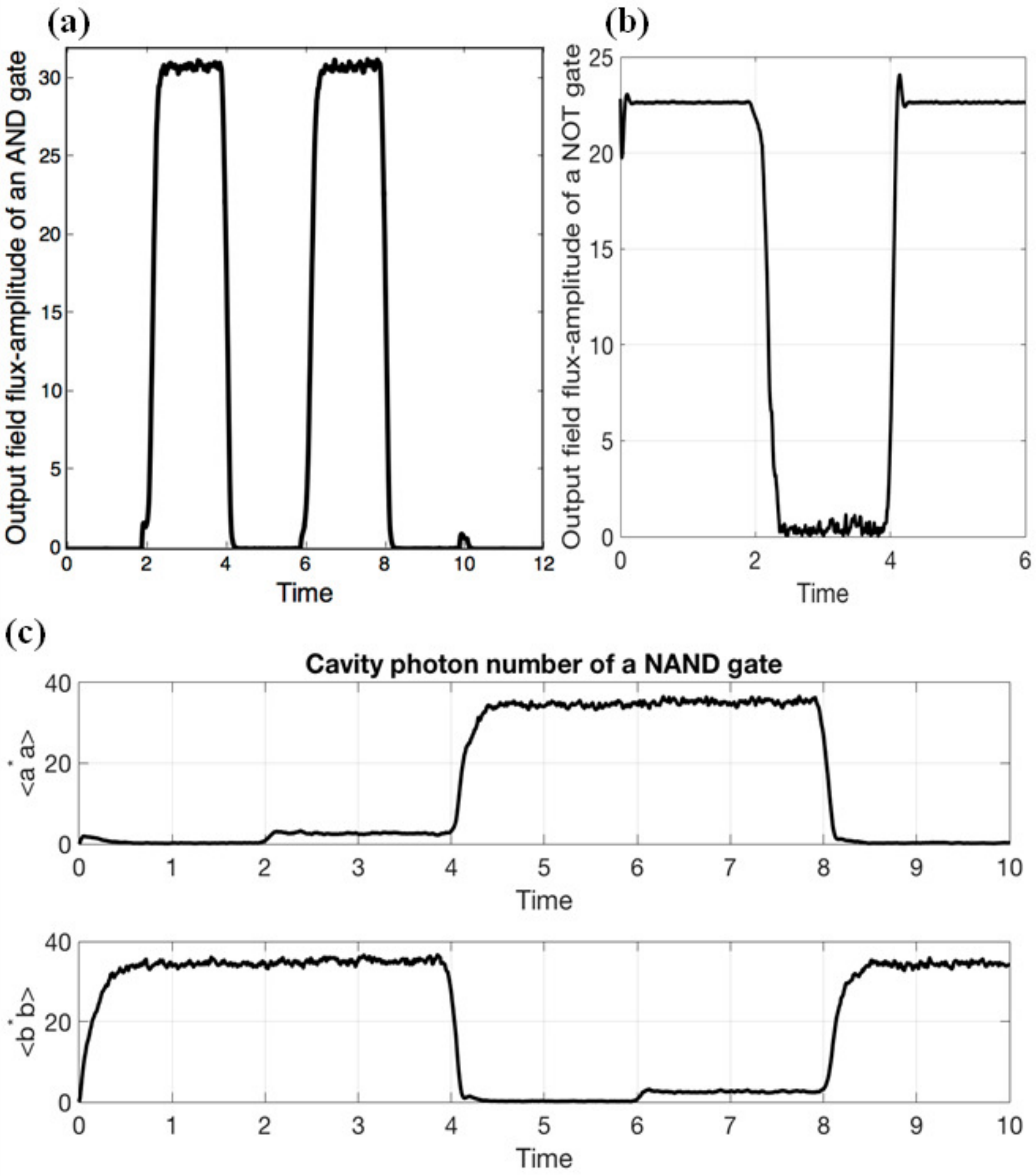}% Here is how to import EPS art
		\caption{\label{fig:gates_full} (a) The magnitude of the mean output field of the full AND gate model as a function of time.  For time 0-2, $\left\langle\xi_1\right\rangle=0, \left\langle\xi_2\right\rangle=0$; 2-4, $\left\langle\xi_1\right\rangle=\alpha, \left\langle\xi_2\right\rangle=\alpha$; 4-6, $\left\langle\xi_1\right\rangle=\alpha, \left\langle\xi_2\right\rangle=0$; 6-8, $\left\langle\xi_1\right\rangle=\alpha, \left\langle\xi_2\right\rangle=\alpha$; 8-10, $\left\langle\xi_1\right\rangle=0, \left\langle\xi_2\right\rangle=\alpha$; 10-12, $\left\langle\xi_1\right\rangle=0, \left\langle\xi_2\right\rangle=0$. 
			(b) The magnitude of the mean output field of the full NOT gate model as a function of time. 
			For time 0-2, $\left\langle\xi\right\rangle=0$; 2-4, $\left\langle\xi\right\rangle=\alpha$; 4-6, $\left\langle\xi\right\rangle=0$.
			(c) Mean cavity photon numbers $\left\langle a^* a \right\rangle$ and $\left\langle b^* b \right\rangle$ of the full NAND latch model as functions of time.
			For time 0-2, $\overline{S}=0, \overline{R}=\alpha$; 2-4, $\overline{S}=\alpha, \overline{R}=\alpha$; 5-6, $\overline{S}=\alpha, \overline{R}=0$; 6-8, $\overline{S}=\alpha, \overline{R}=\alpha$; 8-10, $\overline{S}=0, \overline{R}=\alpha$.
			The time for switching from the low value to the high value of the logic states, and vice-versa, of an input fields is 0.2. }
	\end{center}
\end{figure}

The Kerr nonlinearity can be utilized to realize photonic logic gates, such as an AND gate, a NOT gate and a NAND latch, as  proposed in \cite{Mabuchi2011}. As shown in Figure~\ref{fig:gates}, AND and NOT gates are single cavity systems, while the NAND latch consists of two cavities.  
With $\varphi=1.572$ and $\theta=1.073$, the AND gate has two coherent ingoing fields $\xi_1$ and $\xi_2$ with adjustable amplitudes. The ingoing signals have two logical states HIGH and LOW corresponding to field amplitudes $\alpha=22.6274$ and $0$, separately. The gate has one logical output $\eta_a$ which is HIGH (amplitude is around $31$) when both inputs are HIGH, and LOW (amplitude is 0) in all other cases, see Figure~\ref{fig:gates_full}(a) as an example. 
The NOT gate has three coherent inputs $\xi_\alpha$, $\xi_\beta$ and $\xi_{\beta'}$  with fixed amplitudes $\alpha=22.6274$, $\beta=-34.289-11.909\imath$ and $\beta'=7.833-17.656 \imath$, respectively. It also has a logical input $\xi$ with an adjustable amplitude which is in the HIGH state when $\langle\xi\rangle=\alpha$ and LOW when $\langle\xi\rangle=0$. With  $\theta=0.891$, $\theta'=1.071$ and $\varphi'=2.03$,
the logical output $\eta_n$ is in the HIGH state (amplitude is around $\alpha$) when $\xi$ is LOW and vice-versa.
The NAND latch has two Kerr-cavities with cavity modes $a$ and $b$ connected in a coherent feedback loop. Incoming coherent fields $\xi_{\overline{S}}$ and $\xi_{\overline{R}}$ act as logical input signals with amplitudes $\overline{S}$ and $\overline{R}$ respectively being either $\alpha$ or $0$. Two ingoing signals marked as $\xi_\beta$ are coherent fields with the fixed amplitude $\beta=-34.289-11.909 \imath$. Other numerical system parameters are $\theta=0.891$, $\theta'=0.566$,  $\varphi =2.546$ and $\varphi' =0.158$.
As can be seen in Figure~\ref{fig:gates_full}(c), the NAND latch has three logic states: 
the SET state for $\overline{S}=0$ and $\overline{R}=\alpha$, corresponding to 
%high amplitude (around $\alpha$) of outgoing field $\eta_n$, 
a low average cavity photon number $\left\langle a^* a \right\rangle$ (around $0$) and a high average cavity photon number $\left\langle b^* b \right\rangle$ (around $35$); 
the RESET state for $\overline{S}=\alpha$ and $\overline{R}=0$, corresponding to %
%low amplitude of outgoing field $\eta_n$, 
a high value of $\left\langle a^* a \right\rangle$ and a low value of $\left\langle b^* b \right\rangle$;
and the HOLD state for $\overline{S}=\overline{R}=\alpha$, when the system keeps the previous state.
A  brief introduction to the gates and derivations of their SLH coefficients can be found in Appendix.

We now explore the idea of model order reduction. The main difficulty is that with the chosen parameter values for the Kerr-cavity there is no obvious  limit that one could take. Compare this with the regime $\chi \gg \kappa, \alpha, \Delta$ examined in \cite{Mab12} where a qubit limit of a Kerr-cavity can be obtained as a reduced model by adiabatic elimination (allowing $\chi \rightarrow \infty$). Now, let $\rho_\epsilon$ be the steady-state density matrix of a full Kerr nonlinear cavity model defined on $\mathcal{H} ^{(N)}$ with the driving field amplitude $\epsilon$, and 
let $\rho_\epsilon^{(r)}$ be the corresponding one of a reduced model living in the subspace $\mathcal{H} ^{(d)}$ with $d<N$. 
Following Figure~\ref{fig:gates} and the input amplitude values given above, we see that three critical amplitude values are employed to drive the Kerr nonlinear cavities of the logic gates, which are $\alpha/\sqrt{2} \approx 16$, $\sqrt{2} \alpha \approx 32$ and $0$. 
Thus, we would like to have $\rho_\epsilon$ and $\rho_\epsilon^{(r)}$ to be close for $0 \leq \epsilon \leq \sqrt{2}\alpha$, especially when $\epsilon$ is $\sqrt{2}\alpha$, $\alpha / \sqrt{2}$ and $0$. Inspired by principal components analysis, which is a dimensionality reduction algorithm transforming a data set to a new coordinate system by analyzing the covariance matrix of the data \cite{bJolliffe2002} with applications in pattern recognition \cite{Fuku90} and  signal compression in signal processing \cite{Clarke85,Jain89}, we propose a quasi-principal components approach that reduces the Hilbert space dimension by truncating a particular new basis set $\{T_j\} _{1 \leq j \leq N}$, instead of truncating the conventional Fock state basis. With this in mind, our idea for model reduction is to reduce the dimension of the Hilbert space associated with the density matrices $\rho_\lambda$ and  $\rho_0$ for some $\lambda \in [\alpha / \sqrt{2}, \sqrt{2} \alpha]$. As will be explained in more detail shortly, the qualifier ``quasi'' here is due to the fact that $\rho_0$ and $\rho_{\lambda}$ are in general non-commuting density matrices which cannot be simultaneously diagonalized but only approximately so. The  basis vectors obtained from this process are thus not true principal components associated with the density matrices but only approximate ones.

Suppose that $T$ is a unitary matrix and let $P$ be a $d \times N$ projection matrix onto a proper subspace of $\mathcal{H}^{(N)}$ of dimension $d<N$. We seek a reduced model of a Kerr-cavity with SLH parameters $(S_r, L_r, H_r)$ of the form
\begin{eqnarray}
H_{r} =PT^*HTP^*, ~L_{r} = PT^*LTP^*, ~S_{r} =S=I_2. \label{eq:reduced_cavity}
\end{eqnarray} 
Note that we have $S_{r}=S$, since here $S=I_2$ is a constant matrix for the Kerr-cavity model, rather than a matrix of operators. The question is then how should we choose $T$? If the density matrices $\rho_{0}$ and $\rho_{\lambda}$ commute then they are simultaneously diagonalizable \cite{bHorn1990} and share the same set of mutually orthogonal eigenvectors. We may want to choose $T$ as the simultaneous  eigenvectors corresponding to the largest of the  product of the corresponding eigenvalues of $\rho_0$ and $\rho_{\lambda}$. This is reminiscent of the so-called balanced truncation approach that is commonly employed in control theory for model reduction of linear systems  by identifying $\rho_0$ and $\rho_{\lambda}$ with the controllability and observability Gramian of a linear system \cite{ZDG95}. However, in general, there is no reason to expect that $\rho_0$ and $\rho_{\lambda}$  would commute. Indeed, by computing the commutator $[\rho_\lambda, \rho_0]$ with $\lambda \in \{\alpha/\sqrt{2}, 0.8 \alpha, 0.9\alpha, \alpha, 1.1\alpha,1.2\alpha,1.3\alpha,\sqrt{2} \alpha \}$, we find that $\rho_\lambda$ and $ \rho_l$ do not commute. Therefore, we instead seek to make $\rho_0$ and $\rho_{\lambda}$ approximately simultaneously diagonal. To this end, we propose a method which employs a Jacobi-like algorithm called the  JADE algorithm to approximately simultaneously diagonalize the steady-state matrices $\rho_{\lambda}$ and $\rho_0$ as $\Sigma_j=T^* \rho_j T$ ($j=\{\lambda,0\}$). In the JADE algorithm this is achieved by minimizing a function
$
J( \rho_\lambda, \rho_0, T)  = {\rm off}( T^* \rho_\lambda T) + {\rm off}( T^* \rho_0 T), \nonumber
$
under the constraint $T^*T=I_N$, where $ {\rm off}( A)$ denotes the sum of squared absolute values of all the off-diagonal elements of the matrix $A$ \cite{Cardoso1996,Gerstner1993,Glashoff2013}. With the aid of the Matlab function $\texttt{joint\_diag.m}$ provided in  \cite{Cardoso1996}, which implements the simultaneous approximate diagonalization of complex matrices, we obtain the sum of the approximately diagonalized matrices $\Sigma=\Sigma_\lambda+\Sigma_0$, with the main diagonal elements ordered starting from the top left corner according to their magnitudes, and the corresponding matrix $T$. Consequently, the projection matrix $P$ is simply $P= [\begin{array}{cc} O_{d,N-d} & I_d \end{array}]$. The columns of $T$ can be viewed as quasi-principal components or vectors associated with $\rho_{0}$ and $\rho_{\lambda}$  that become genuine principal components when $\rho_0$ and $\rho_{\lambda}$ commute.

We now examine the JADE-based model reduction method for the steady-state outputs of a Kerr nonlinear cavity, with various choices of $\lambda$. We denote the corresponding reduced model for a particular choice of $\lambda$ as $G_\lambda^{(r)}$.
As shown in Figure~\ref{fig:cav_outputs}, with $d=15$ the reduced model $G_\alpha^{(r)}$ matches the full model well for driving field amplitudes in the range $0 \leq \epsilon \leq \sqrt{2}\alpha$. The reduced models with $\lambda<\alpha$ have a good match when $\epsilon$ is under the nonlinearity threshold $\epsilon< 26$ and poor performances for higher driving strength, while the models with $\lambda>\alpha$ do not perform well for low driving amplitudes but match well for the high amplitudes above the threshold. Therefore, from now on we set $\lambda =\alpha$ and take $G_\alpha^{(r)}$ as the reduced cavity model which will be applied in the logic gate models later. 

\begin{figure}[htbp]
	\begin{center}
		\includegraphics[scale=0.22]{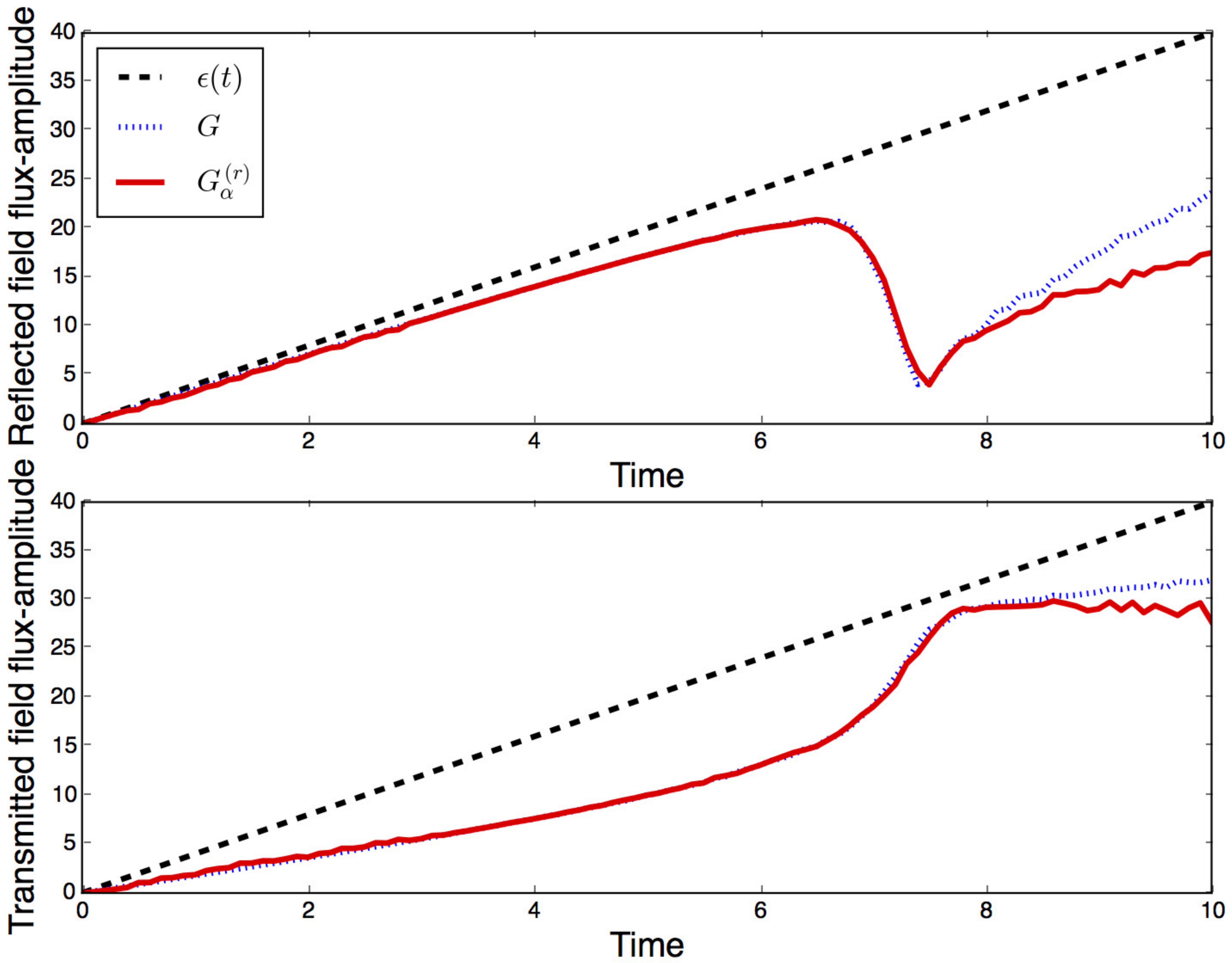}% Here is how to import EPS art
		\caption{\label{fig:cav_outputs_time} Comparison of time-varying reflected (top) and transmitted (bottom) magnitudes of the mean output fields of the full cavity model $G$ and the JADE-based reduced model $G_\alpha^{(r)}$, with $d=15$ and the time-varying cavity driving field $\epsilon(t)= 4t$.}
	\end{center}
\end{figure}
Figure~\ref{fig:cav_outputs_time} compares the time-varying magnitudes of the mean output fields of the full model $G$ and the reduced model $G_\alpha^{(r)}$, with a time-varying cavity driving field $\epsilon(t)=4t$ ($0 \leq t \leq 10$). 
We note that all the numerical simulations of dynamics of the cavity optics models in this paper are implemented by a quantum Monte Carlo solver with 100 trajectories \cite{Tan1999,Johansson2012}. The simulations were run on an Apple Macbook Pro laptop configured with a 2.5GHz quad-core Intel Core i7 processor, 16 GB of memory and 500 GB flash storage\footnote{We have employed either the  Quantum Optics Toolbox in Python (QuTiP) \cite{Johansson2012} or Quantum Optics Toolbox for MATLAB (qotoolbox) \cite{Tan1999} to simulate cavity nonlinear optics, depending on which toolbox is more convenient to handle time-dependent Hamiltonians with the time-varying cavity driving fields of a certain quantum system. Explicitly, the time-evolution of a Kerr nonlinear cavity and an AND gate were simulated by QuTiP and the rest of simulations were implemented with qotoolbox.}.  Moreover, for all simulations in this work, the full model of the Kerr-cavity is initialized in the Fock vacuum $|0\rangle$  while reduced models were initialized in the projection of $|0\rangle$ onto the Hilbert space of the reduced models.
It is shown that the magnitude of the mean output field of the reduced model closely follows the full model one  for $0\leq \epsilon \leq \sqrt{2}\alpha$ for amplitudes up to $\approx 32$ (i.e., up to $t \approx 8$), and in this range agrees with the steady-state plots shown in  Figure~\ref{fig:cav_outputs}.  Although the reduced model mean outputs no longer  follow the full model mean outputs as the magnitude is increased beyond 32, we emphasize that this high amplitude region is not of interest for us since we are primarily interested in the region of weak excitations with low mean photon numbers.
Furthermore, the reduced model is more efficient as it requires only 276.50 seconds of simulation time, while the full model consumes 320.49 seconds.

\begin{figure}[htbp]
	\begin{center}
	\includegraphics[scale=0.5]{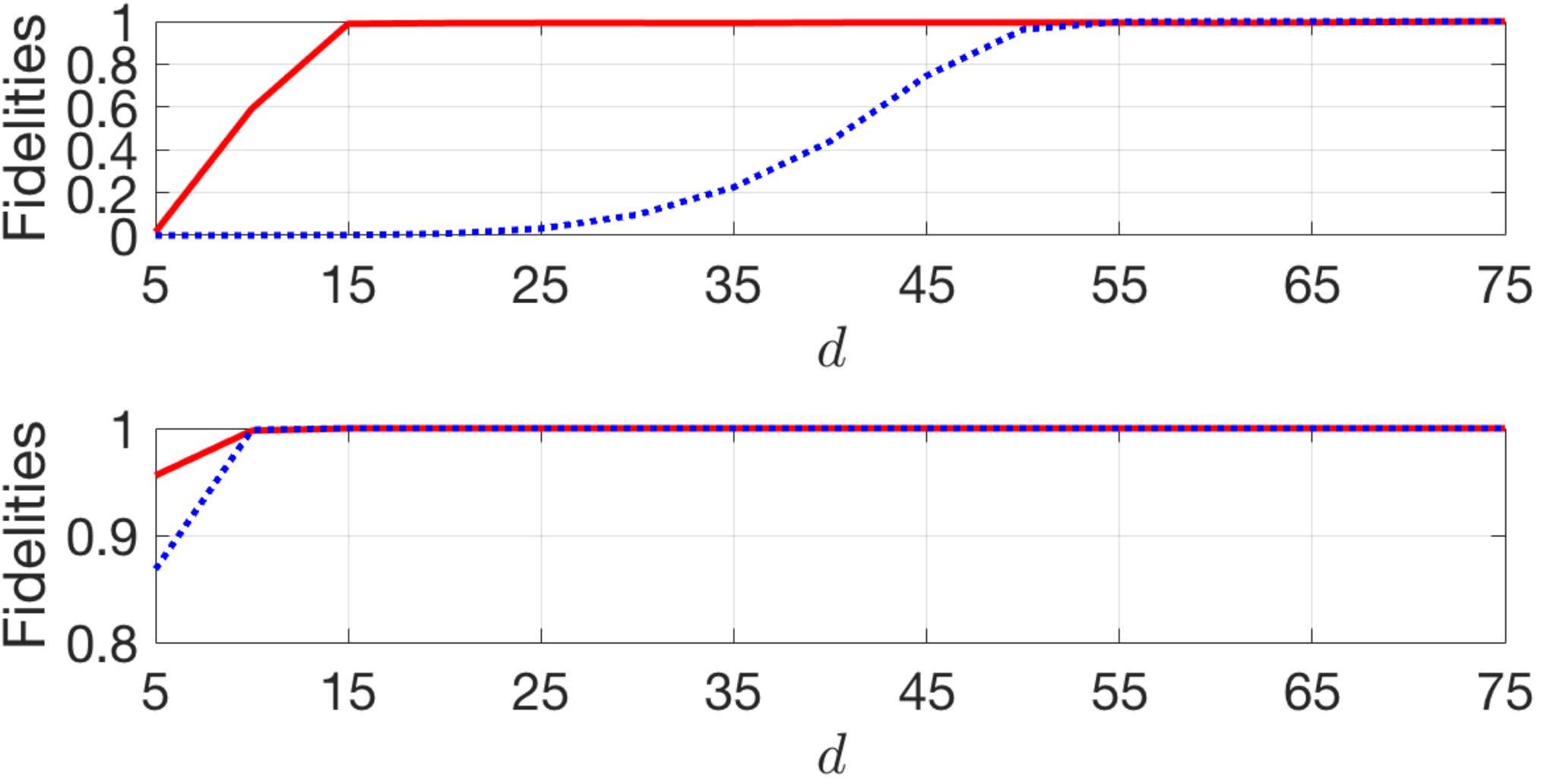}% Here is how to import EPS art
	\caption{\label{fig:fidelities} Fidelity comparison between $F(\rho(\epsilon), \rho_{Jr\rightarrow F}(\epsilon))$ (red solid line) and $F(\rho(\epsilon), \rho_{Fr\rightarrow F}(\epsilon))$ (blue dotted line), with $d=\{5,10, 15,20, 25, 30, 35, 40, 45, 50, 55, 60, 65, 70, 75\}$ and driving fields strength $\epsilon=\sqrt{2}\alpha$ (top) and $\epsilon=\alpha/\sqrt{2}$ (bottom).}
	\end{center}
\end{figure}
To illustrate the merits of the JADE-based model reduction method, we now compare it with the well-known Fock-state-based reduction method that truncates the Fock states corresponding to large photon numbers. The resulting $T$ and $P$ matrices of the Fock-state-based method are $T_{F}=I_N$ and $P_{F} = [\begin{array}{cc} I_d & O_{d,N-d} \end{array}]$.
We look at fidelities between steady states of the full model and the two reduced models.  
Fidelity is a measure of distance between two quantum states, say $\rho$ and $\sigma$, which is defined as
$
F(\rho, \sigma) = {\rm tr} \sqrt{\sqrt{\rho} ~ \sigma \sqrt{\rho}}.
$
Let the steady states of the full model, JADE-based reduced model, and truncated Fock state-based reduced model (of dimension $d<N$) ) be $\rho(\epsilon)$, $\rho_{Jr}(\epsilon)$ and  $\rho_{Fr}(\epsilon)$, respectively, as functions of the driving strength $\epsilon$. To calculate fidelity, we need to convert the representations $\rho_{Jr}(\epsilon)$ in the basis $\{T_j\} _{1 \leq j \leq N}$ and  $\rho_{Fr}(\epsilon)$  with respect to the reduced Fock-state basis to the ones in the same Fock state basis of $\rho(\epsilon)$. The resulting new representations are
$$
\rho_{Jr\rightarrow F}(\epsilon)=T \left[\begin{array}{cc} O_{n-d,d} & O_{n-d,n-d}\\O_{d,n-d} &\rho_{Jr}(\epsilon) \end{array}\right]T^*
$$
and
$$
\rho^{ss}_{Fr\rightarrow F}(\epsilon) =\left[\begin{array}{cc} \rho_{Fr}(\epsilon) & O_{d,n-d}\\O_{n-d,d} & O_{n-d,n-d}\end{array}\right],
$$
for  $\rho_{Jr}(\epsilon)$ and $\rho_{Fr}(\epsilon)$, respectively. Figure~\ref{fig:fidelities} illustrates the fidelities $F(\rho(\epsilon), \rho_{Jr\rightarrow F}(\epsilon))$ (red solid line) and $F(\rho(\epsilon), \rho_{Fr\rightarrow F}(\epsilon))$ (blue dotted line), with the driving strength $\epsilon=\sqrt{2}\alpha$ (top) and $\epsilon=\alpha/\sqrt{2}$ (bottom), with respect to various values of the reduced model dimension $d$.
We see that the steady state of the JADE-based reduced model becomes very close to the one of the full model ($F(\rho(\epsilon), \rho_{Jr\rightarrow F}(\epsilon)) \uparrow 1 $) when $d \geq 15$ for $\epsilon=\sqrt{2}\alpha$ and $d\geq10$ for $\epsilon=\alpha/\sqrt{2}$.
On the other hand, to have a high fidelity close to $1$, the Fock-state-based reduced model has to keep a higher dimension (at least $55$) when the cavity driving strength is strong, while the dimension can be reduced to $10$ when the driving field amplitude is low. 
Thus, compared to the Fock-state based method, the JADE-based model reduces the model order to a lower dimension while keeping a high fidelity for both driving amplitudes.

\begin{remark}
Note that it is possible to consider approximate simultaneous diagonalization of more than two density matrices $\rho_0$ and $\rho_{\alpha}$. For instance, one may try to apply the JADE algorithm to the three density matrices $\rho_{0}$, $\rho_{\alpha/\sqrt{2}}$ and $\rho_{\sqrt{2}\alpha}$. Although not detailed here, the approximate simultaneous diagonalization with these three density leads to a reduced model of Hilbert space dimension 22 which tracks the magnitudes of the mean output fields of interest more accurately than the reduced model presented here. However, this higher dimensional reduced model takes longer to simulate than the full model despite its lower dimension, especially when inserted in a logic gate configuration that will  be discussed in the next section. We note that no attempt to optimize the numerical simulation of this 22 dimension reduced model was undertaken.
\end{remark}
%%%%%%%%%%%%%%%%%%%%%%%%%%%%%%%%%%%%%%%%%%%%%%%%%%%%%%%%%%%%%%
%%%%%%%%%%%%%%%%%%%%%%%%%%%%%%%%%%%%%%%%%%%%%%%%%%%%%%%%%%%%%%
\section{\label{sec:gates} Model reduction of cavity nonlinear optical models}
In this section we present three model reduction examples for an AND gate, a NOT gate and a NAND latch built from a single Kerr nonlinear cavity (the AND and NOT gates) or two Kerr nonlinear cavites (the NAND latch) as described in Section~\ref{sec:kerr_cavity}, which are denoted by $G_A, G_N$ and $G_L$ separately. Their reduced models, denoted by $G_{A}^{(r)}$, $G_{N}^{(r)}$ and $G_{L}^{(r)}$, are obtained by replacing each full Kerr-cavity model in the gates and latch with the JADE-based reduced cavity model $G_\alpha^{(r)}$ with $d=15$. The dimensions of the Hilbert spaces of the reduced models are $d$,  $d$ and $d^2$, respectively. SLH coefficients of the reduced models are derived in Appendix.

\begin{figure}[htbp]
	\begin{center}
		\includegraphics[scale=0.65]{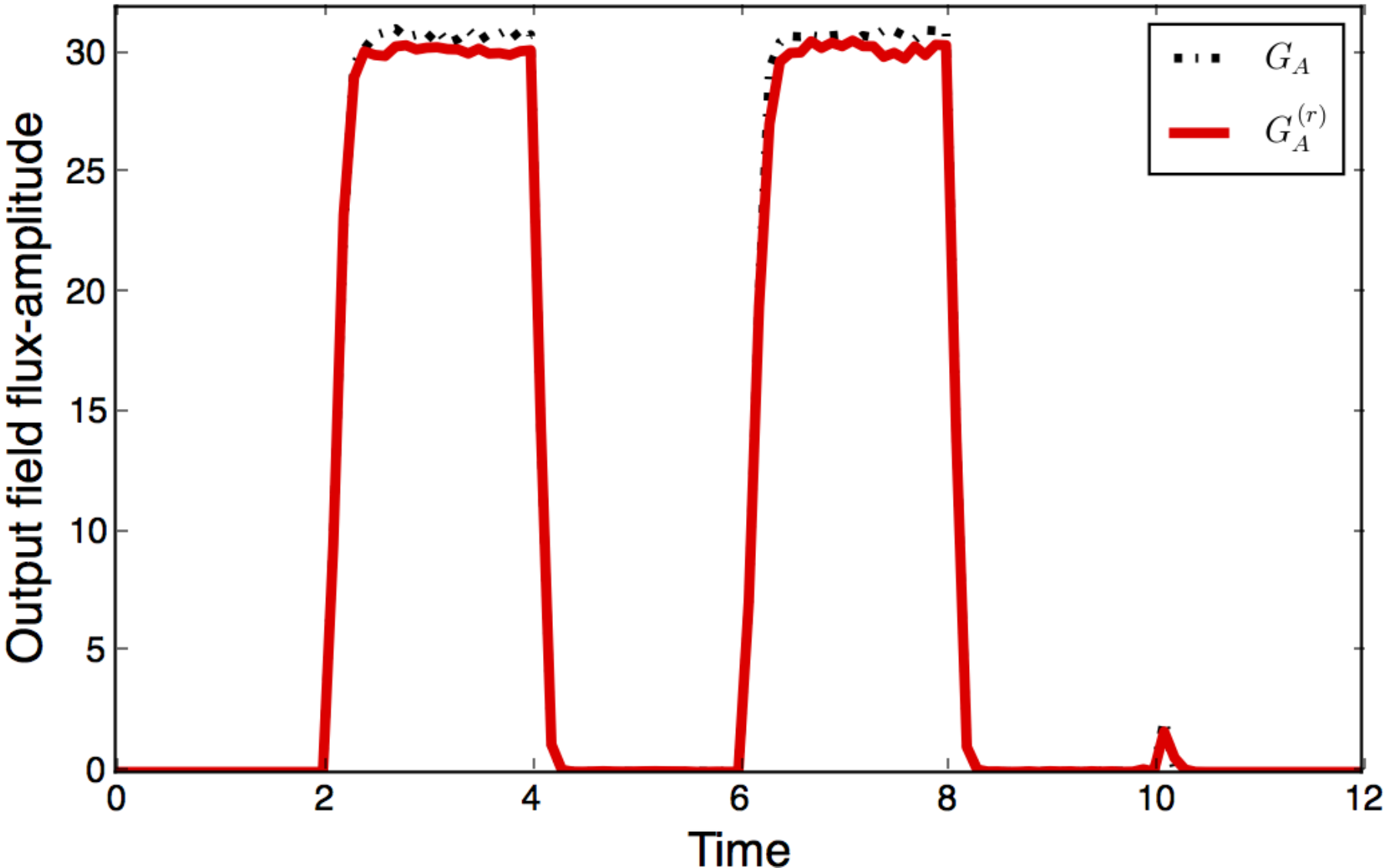}% Here is how to import EPS art
		\caption{\label{fig:and_gate_compare} Comparison of the magnitudes of the mean output fields of the full AND gate model and the JADE-based reduced model with $d=15$ as functions of time.  For time 0-2, $\left\langle\xi_1\right\rangle=0, \left\langle\xi_2\right\rangle=0$; 2-4, $\left\langle\xi_1\right\rangle=\alpha, \left\langle\xi_2\right\rangle=\alpha$; 4-6, $\left\langle\xi_1\right\rangle=\alpha, \left\langle\xi_2\right\rangle=0$; 6-8, $\left\langle\xi_1\right\rangle=\alpha, \left\langle\xi_2\right\rangle=\alpha$; 8-10, $\left\langle\xi_1\right\rangle=0, \left\langle\xi_2\right\rangle=\alpha$; 10-12, $\left\langle\xi_1\right\rangle=0, \left\langle\xi_2\right\rangle=0$. 
			The time for switching from the low value to the high value of the logic states, and vice-versa,  of the input fields is 0.2.}
	\end{center}
\end{figure}

\begin{figure}[htbp]
	\begin{center}
		\includegraphics[scale=0.62]{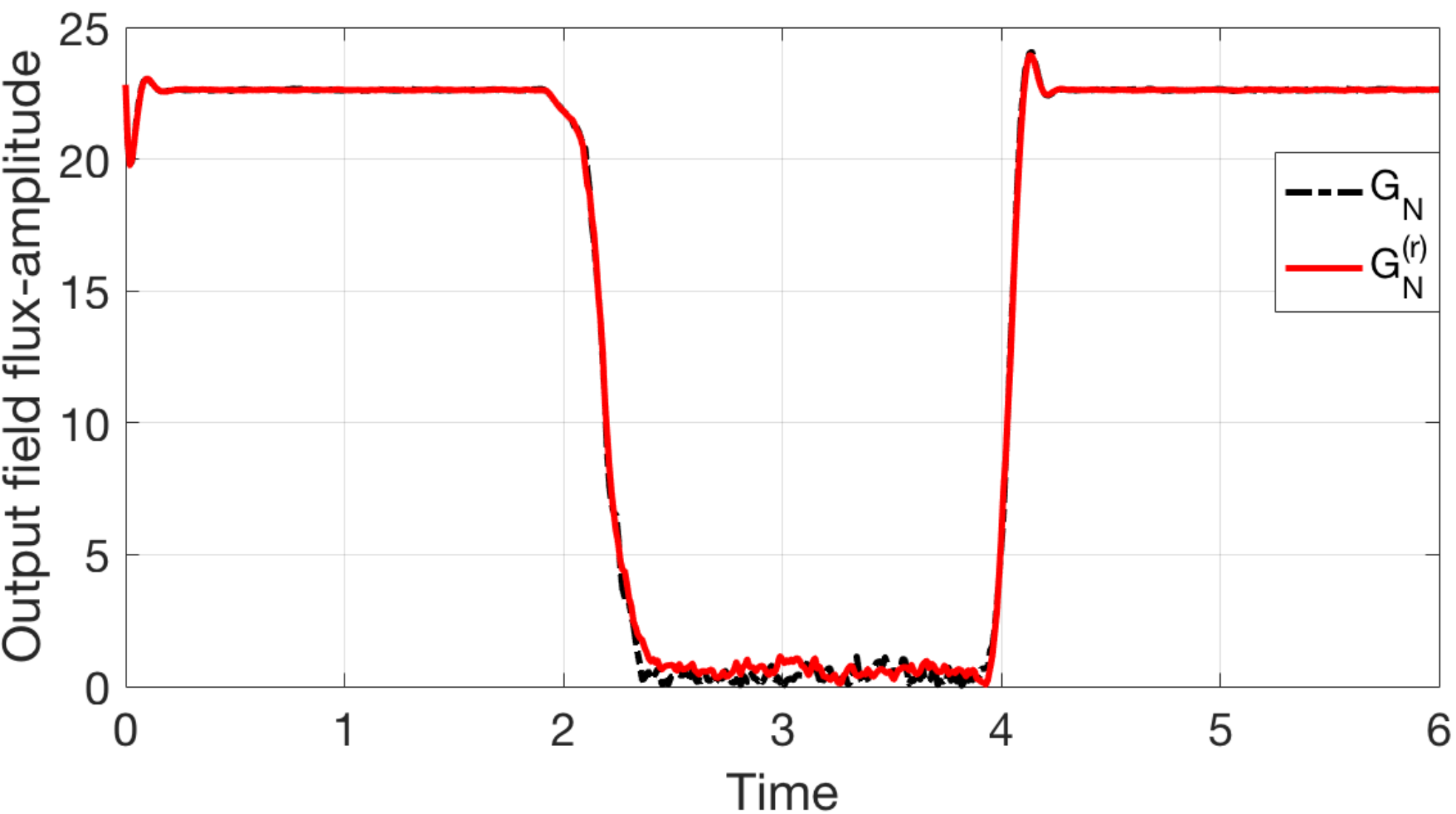}% Here is how to import EPS art
		\caption{\label{fig:not_gate_compare} Comparison of the magnitudes of the mean output fields of the full NOT gate model and the JADE-based reduced model with $d=15$ as functions of time. 
			For time 0-2, $\left\langle\xi\right\rangle=0$; 2-4, $\left\langle\xi\right\rangle=\alpha$; 4-6, $\left\langle\xi\right\rangle=0$. The time for switching from the low value to the high  value of the logic states, and vice-versa,  of the input field is 0.2.}
	\end{center}
\end{figure}

We firstly look at model reduction of the two single-cavity gates. 
Figure~\ref{fig:and_gate_compare} and Figure~\ref{fig:not_gate_compare} present the magnitudes of the mean output fields of the original and JADE-based reduced models of AND and NOT gates separately, with inputs in various logical states. 
We see quantitatively that the outputs of the reduced models well-match the original systems. Moreover, the total simulation time of the full and reduced AND gate models are 344.86 and 327.57 seconds, separately; the total run time of the full and reduced NOT gate models are 77.33 and 62.68 seconds, respectively.

\begin{figure}[htbp]
	\begin{center}
	\includegraphics[scale=0.63]{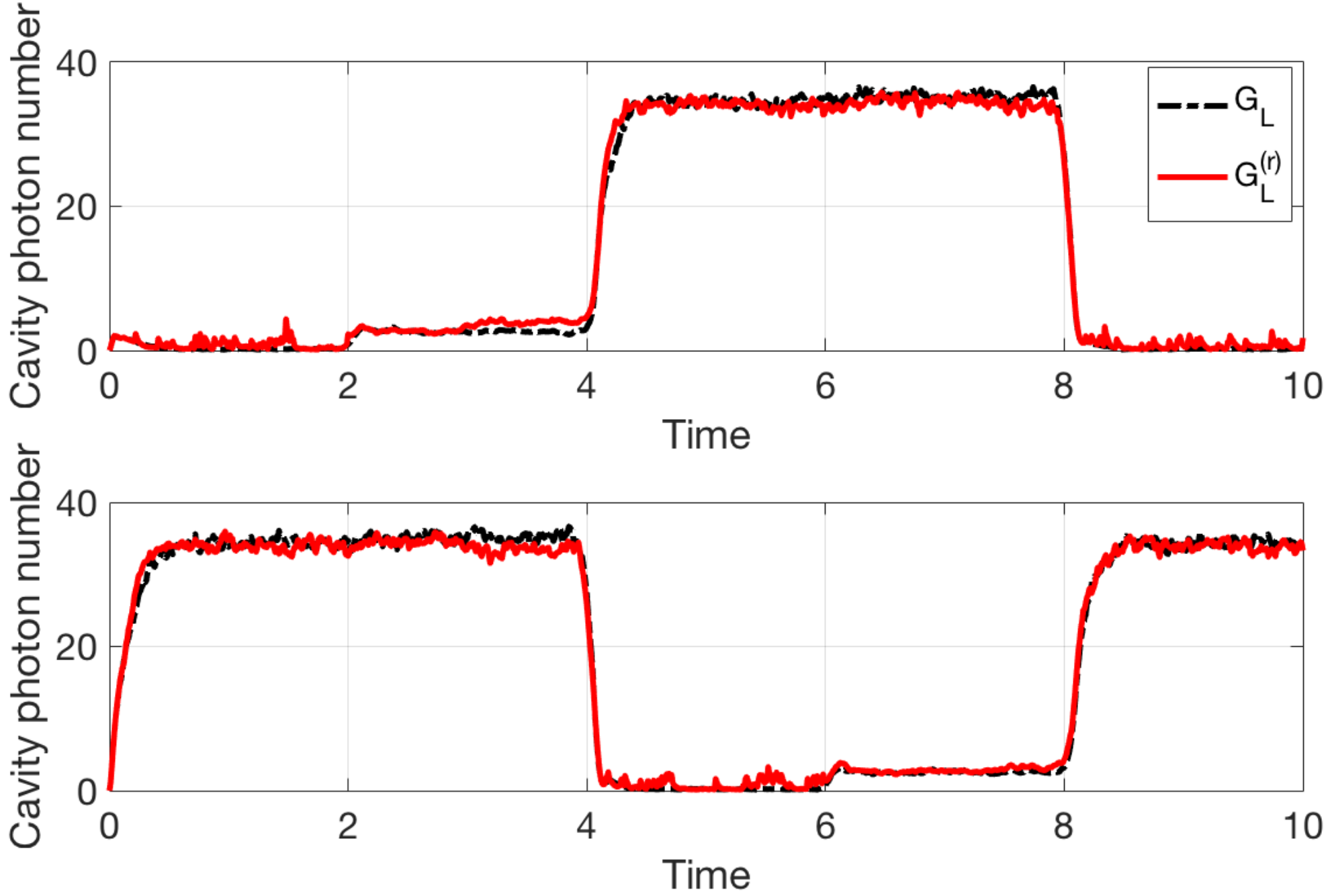}% Here is how to import EPS art
	\caption{\label{fig:nand_gate_compare} Comparison of mean cavity photon numbers $\left\langle a^* a \right\rangle$ and $\left\langle b^* b \right\rangle$ of the original NAND latch model and the JADE-based reduced model with $d=15$ as functions of time.
		For time 0-2, $\overline{S}=0, \overline{R}=\alpha$; 2-4, $\overline{S}=\alpha, \overline{R}=\alpha$; 5-6, $\overline{S}=\alpha, \overline{R}=0$; 6-8, $\overline{S}=\alpha, \overline{R}=\alpha$; 8-10, $\overline{S}=0, \overline{R}=\alpha$. The time for switching from the low value to the high of the logic states, and vice-versa,  of the input fields is 0.2.}
	\end{center}
\end{figure}

Finally, we investigate model reduction of the NAND latch. Figure~\ref{fig:nand_gate_compare} compares cavity average photon numbers $\left\langle a^* a \right\rangle$ and $\left\langle b^* b \right\rangle$ of the original and JADE-based reduced models.
The total simulation time of a full latch model is 25896.88 seconds, while the run time for the reduced model is 11605.42 seconds. Therefore, the reduced model not only yields a good match, but also saves more than half of the simulation time for this two-cavity model.

\section{\label{sec:conclu} Conclusion}
In this paper, we have proposed a model reduction approach for Kerr nonlinear optical cavities, based on the idea of quasi-principal  components and exploiting approximate simultaneous diagonalization of density operators using an algorithm known as the JADE algorithm. We first employed the proposed method to obtain a reduced QSDE model for a single Kerr nonlinear cavity from two steady-state matrices $\rho_\lambda$ and $\rho_0$ corresponding to the driving field amplitudes $\lambda$ and $0$,  respectively, with $\alpha / \sqrt{2} \leq \lambda \leq \sqrt{2}\alpha$. We found a reduced model of dimension 15 with  $\lambda=\alpha$ that can closely match the steady-state and time-varying magnitudes of the mean output fields of a full model based on truncation to 75 Fock states of the oscillator Hilbert space. After that, the reduced models of  a Kerr-cavity-based AND gate, NOT gate, and a NAND latch were obtained by replacing each full cavity model in the gates or latch with the reduced cavity model. Our simulation results illustrate that the gates with the reduced model match the ones with the full model well. Furthermore, the reduced models consume less simulation time, and save more than half the simulation time in the simulation of the NAND latch that consists of the interconnection of two Kerr nonlinear cavities. These savings were obtained without any attempts to optimize the numerical simulations with the reduced model.

Although developing model reduction techniques for individual nonlinear input-output open quantum systems such as a single Kerr nonlinear optical cavity  is an important topic in its own right, ultimately one would like to use the reduced quantum models to simulate medium to  large scale photonic logic circuits that consist of  multiple Kerr-cavities. Despite the reduced model having a Hilbert space of dimension 15 and is a  fully quantum model, when there are multiple Kerr-cavities the dimension of a circuit would still grow exponentially quickly. Thus further research is needed to explore whether there are further features of the 15 dimension reduced model that combined together with new efficient simulation techniques that will allow simulation for large logic circuits made of several Kerr cavities and other nonlinear optical components. Since the quantum models here are actually employed to simulate a classical logic circuit, there is reason to believe that the individual Kerr cavities in the circuit are likely to be weakly entangled, a scenario which can potentially be amenable to efficient numerical quantum simulation techniques.  For instance, recent efficient numerical techniques for simulating open quantum systems such as matrix product operator techniques may be of interest for this purpose, see, e.g., \cite{VGRC04,CPHJP10,CCB15}. These are topics to be pursued in future investigations.

%%%%%%%%%%%%%%%%%%%%%%%%%%%%%%%%%%%%%%%%%%%%%%%%%%%%%%%%%%%%%%%%%%%%%%%%%%%%%%%%%%%%%%%%%%%%%%%%%%%%%%%%%%%%%%%%%%%%%%%%%%%%%%%%%%%%%%%%%%%%%%%%%%%%%%%%%
%%%%%%%%%%%%%%%%%%%%%%%%%%%%%%%%%%%%%%%%%%%%%%%%%%%%%%%%%%%%%%%%%%%%%%%%%%%%%%%%%%%%%%%%%%%%%%%%%%%%%%%%%%%%%%%%%%%%%%%%%%%%%%%%%%%%%%%%%%%%%%%%%%%%%%%%%
%%%%%%%%%%%%%%%%%%%%%%%%%%%%%%%%%%%%%%%%%%%%%%%%%%%%%%%%%%%%%%%%%%%%%%%%%%%%%%%%%%%%%%%%%%%%%%%%%%%%%%%%%%%%%%%%%%%%%%%%%%%%%%%%%%%%%%%%%%%%%%%%%%%%%%%%%
%%%%%%%%%%%%%%%%%%%%%%%%%%%%%%%%%%%%%%%%%%%%%%%%%%%%%%%%%%%%%%%%%%%%%%%%%%%%%%%%%%%%%%%%%%%%%%%%%%%%%%%%%%%%%%%%%%%%%%%%%%%%%%%%%%%%%%%%%%%%%%%%%%%%%%%%%
%%%%%%%%%%%%%%%%%%%%%%%%%%%%%%%%%%%%%%%%%%%%%%%%%%%%%%%%%%%%%%%%%%%%%%%%%%%%%%%%%%%%%%%%%%%%%%%%%%%%%%%%%%%%%%%%%%%%%%%%%%%%%%%%%%%%%%%%%%%%%%%%%%%%%%%%%
%%%%%%%%%%%%%%%%%%%%%%%%%%%%%%%%%%%%%%%%%%%%%%%%%%%%%%%%%%%%%%%%%%%%%%%%%%%%%%%%%%%%%%%%%%%%%%%%%%%%%%%%%%%%%%%%%%%%%%%%%%%%%%%%%%%%%%%%%%%%%%%%%%%%%%%%%
%%%%%%%%%%%%%%%%%%%%%%%%%%%%%%%%%%%%%%%%%%%%%%%%%%%%%%%%%%%%%%%%%%%%%%%%%%%%%%%%%%%%%%%%%%%%%%%%%%%%%%%%%%%%%%%%%%%%%%%%%%%%%%%%%%%%%%%%%%%%%%%%%%%%%%%%%
%%%%%%%%%%%%%%%%%%%%%%%%%%%%%%%%%%%%%%%%%%%%%%%%%%%%%%%%%%%%%%%%%%%%%%%%%%%%%%%%%%%%%%%%%%%%%%%%%%%%%%%%%%%%%%%%%%%%%%%%%%%%%%%%%%%%%%%%%%%%%%%%%%%%%%%%%
\newpage
\section*{Appendix}
\appendix
This appendix first gives a brief overview of the Hudson-Parthasarathy quantum stochastic differential equation (QSDE), SLH coefficients of an (input-output) open Makov quantum system, and formulae of operations such as concatenation product, series product and feedback on Markovian quantum subsystems. The details can be found in \cite{HP84,KRP92,Mey95,Gough2009,Gough2009math}. 
Afterwards, we present SLH parameters and master equations for cavity nonlinear optics of interest. Finally, we give SLH models for reduced systems of the cavity nonlinear optics.
\section{\label{sec:dynamics} Mathematical description of input-output open Markov quantum systems}
Consider an open Markov quantum system $G=(S, L, H)$ with $n$ ingoing fields and $n$ outgoing fields. 
%The system is characterized by a triple $G=(S, L, H)$ refereed as `SLH', where $H$ is a Hermitian operator specifying the self-energy of the system, $S$ is a unitary scattering matrix with operator elements $\{S_{ij}\}$, $L$ is a coupling operator vector with entries $\{L_j\}$  ($1\leq i, j \leq n$). 
%The operators involved in the triple are defined on an Hilbert space $\mathcal{H}$, refereed as the initial space.
We assume the incoming fields are boson fields described by field annihilation operators $\xi_t^{j}$ and field creation operators ${\xi_t^{j}}^*$ ($1 \leq j \leq n$). 
Recall that the field operators satisfy the singular white noise commutation relations $\left[\xi_t^i, {\xi_s^j}^*\right]=\delta_{ij}\delta(t-s)$ and $\left[\xi_t^i, \xi_s^j\right]=0$. Defining the annihilation processes $A_t^j=\int_{0}^{t} \xi_s^j ds$, creation processes ${A_t^j}^*=\int_{0}^{t} {\xi_s^j }^*ds$ and gauge processes $\Lambda_t^{jk}=\int_{0}^{t}{\xi_s^j }^*\xi_s^k ds$ are quantum stochastic processes whose forward pointing differentials in a coherent state of the field satisfy the  quantum It\={o} product rules shown in Table~\ref{tb:ito},
with $dA_t^k=A_{t+dt}^k-A_t^k$, $d{A_t^k}^* ={A_{t+dt}^k}^*-{A_t^k}^*$ and $d \Lambda_{t}^{kl}=\Lambda_{t+dt}^{kl} -\Lambda_t^{kl} $.
\begin{table}[h!]
	\caption{Quantum It\^{o} rules table} % title of Table
	\centering % used for centering table
	\begin{tabular}{c| c c c}
		\hline
		&$dA_t^k$&$d{A_t^k}^* $&$d \Lambda_t^{kl} $\\ [0.5ex] % inserts table
		\hline
		$dA_t^j $& 0 & $\delta_{jk}dt$& $\delta_{jk}dA_t^l$\\ % inserting body of the table
		$d{A_t^j}^*$  & 0 & 0 & 0 \\
		$d \Lambda_t^{ij} $ & 0& ${\delta_{jk}dA_t^i}^*$ & $\delta_{jk}d\Lambda_t^{il}$ \\
		$dt$ &0&0&0\\ [1ex] % [1ex] adds vertical space
		\hline
	\end{tabular}
	\label{tb:ito} % is used to refer this table in the text
\end{table}

The joint dynamics of the system and the bosonic field it is coupled to is given by the right Hudson-Parthasarahty quantum stochastic differential equation (QSDE)
\begin{eqnarray}
dU_{t} = \left(\sum_{j,k=1}^{n} \left(S_{jk}-\delta_{jk}\right) d\Lambda_{t}^{jk} + dA_t^* L - L^*SdA_t  -\left(\imath H + \frac{1}{2}L^*L \right)dt\right)U_t; ~U_0=I,
\end{eqnarray}
where $U_t$ is a unitary adapted process.
%%%%%%%%%%%%%%%%%%%%%%%%%%%%%%%%%%%%%%%%%%%%%%%%%%%%%%%%%%%%%%%%%%%%%%%%%%%%%%%%%%%%%%%%%%%%%%%%%%%%%%%%%%%%%%%%%%%%%%%%%%%%%%%%%%%%%%%%%%%%%%%%%%%%%%%%%

The time evolution in the Heisenberg picture of a bounded operator $X$ defined on the initial space $\mathcal{H}$ of the system is $X_t=U_t^* X_0 U_t$. Given the initial state of a system represented by a density operator $\rho_0$, the quantum expectation of $X_t$ is $\left\langle X_t \right\rangle  ={\rm Tr}  \left[X_t \rho_0 \right]$. 
Following the quantum It\={o} product table and the quantum It\={o} rule for the differential of the product of two adapted process $U$ and $V$ satisfying a QSDE,  $d(UV)=dUV+UdV+dUdV$, we have
\begin{eqnarray}
&& \left\langle dX_t\right\rangle= \left\langle d\left(U_t^* X_0 U_t  \right)  \right\rangle \nonumber\\
&& =\left\langle dU_t^* X_0 U_t+U_t^* X_0 dU_t +dU_t^* X_0 dU_t  \right\rangle  \nonumber\\
&& ={\rm Tr} \left[\left(U_t^* \left(\imath H - \frac{1}{2}L^*L \right) X_0  U_t - U_t^* X_0 \left(\imath H + \frac{1}{2}L^*L \right) U_t + \sum_{j=1}^{n} U_t^* L_j^* X_0 L_j U_t  \right) dt\rho_0 \right] \nonumber\\
&& ={\rm Tr} \left[ X_0 \left(\rho_t\left(\imath H - \frac{1}{2}L^*L \right)  - \left(\imath H + \frac{1}{2}L^*L \right) \rho_t  + \sum_{j=1}^{n}  L_j \rho_t L_j^*  \right)  dt\right] \nonumber\\
&& ={\rm Tr} \left[ X_0 \left( -\imath \left[H, \rho_t \right]  + \sum_{j=1}^{n}L_j \rho_t L_j^* - \frac{1}{2}\rho_t L^*L   - \frac{1}{2}L^*L \rho_t \right)  dt \right]. \nonumber
\end{eqnarray}
In physics literature, the expectation of an operato can be expressed in the Schr\"{o}dinger picture such that $\left\langle X_t \right\rangle  ={\rm Tr}  \left[X_0 \rho_t \right]$. 
Comparing the above equation gives
\begin{eqnarray}
\left\langle dX_t\right\rangle={\rm Tr} \left[ X_0  d\rho_t \right], 
\end{eqnarray}
and  the master equation that governs the time evolution of the system state as
\begin{eqnarray}
\frac{d\rho_{t}}{dt}=-\imath \left[H, \rho_t \right]  + \sum_{j=1}^{n}L_j \rho_t L_j^* - \frac{1}{2} \{L^*L,\rho_t\}. \label{eq:master_equation}
\end{eqnarray}

%%%%%%%%%%%%%%%%%%%%%%%%%%%%%%%%%%%%%%%%%%%%%%%%%%%%%%%%%%%%%%%%%%%%%%%%%%%%%%%%%%%%%%%%%%%%%%%%%%%%%%%%%%%%%%%%%%%%%%%%%%%%%%%%%%%%%%%%%%%%%%%%%%%%%%%%%
%%%%%%%%%%%%%%%%%%%%%%%%%%%%%%%%%%%%%%%%%%%%%%%%%%%%%%%%%%%%%%%%%%%%%%%%%%%%%%%%%%%%%%%%%%%%%%%%%%%%%%%%%%%%%%%%%%%%%%%%%%%%%%%%%%%%%%%%%%%%%%%%%%%%%%%%%
\section{\label{sec:SLH}  SLH formalism for arbitrary quantum networks}
\begin{figure*}[htbp]
	\includegraphics[scale=0.5]{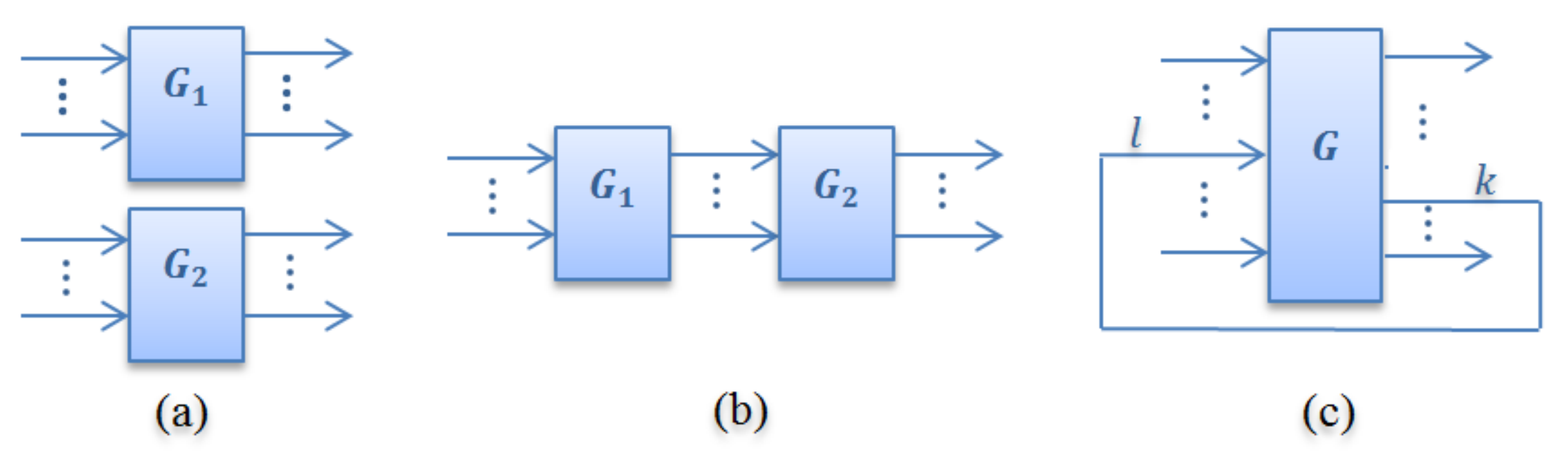}% Here is how to import EPS art
	\caption{\label{fig:S_operations} Algebraic operations of quantum optical components. (a) Concatenation product $G_1 \boxplus G_2$; (b) series product $G_2 \lhd G_1$; (c) feedback $\left[G \right]_{k \rightarrow l}$.}
\end{figure*}

In general, the SLH triple of an arbitrary quantum network can be constructed from triples of its elementary quantum components, such as coherent displacements, cavities, beam splitters and phase shifters, via three algebraic operations: concatenation product, series product and feedback \cite{Gough2009, Gough2009math}.

\subsection{\label{sec:concatenation} Concatenation product}
Given two systems $G_1= (S_1, L_1, H_1)$ with $n_1$ ingoing fields and $G_2= (S_2, L_2, H_2)$ with $n_2$ inputs, the concatenation product of the systems (shown as figure~\ref{fig:S_operations}(a)) is a joint system interacting with $n_1+n_2$ input fields, without any scattering process between $G_1$ and $G_2$, which is of the form
\begin{eqnarray}
G_1 \boxplus G_2 =\left(
\left( \begin{array}{cc} S_1 & O \\ O & S_2 \end{array}\right), 
\left( \begin{array}{c} L_1\\ L_2 \end{array}\right), 
H_1+H_2
\right). \label{eq:concatenation}
\end{eqnarray}

\subsection{\label{sec:series} Series product}
Given $n_1=n_2$, the series product
\begin{eqnarray}
G_2 \lhd G_1 =\left(
S_2S_1, 
L_2+S_2L_1, 
H_1+H_2+ {\rm Im}( L_2^* S_2 L_1)
\right) \label{eq:series} \nonumber\\
\end{eqnarray}
describes a system where the output of its subsystem $G_1$ are fed into the corresponding input ports of $G_2$, see figure~\ref{fig:S_operations}(b).

\subsection{\label{sec:feedback} Feedback}
Consider a system $G= (S, L, H)$ whose outgoing signal of the $k$th output channel is fed back into the $l$th input port, as shown in  figure~\ref{fig:S_operations}(c). The resulting new system $\left[G \right]_{k \rightarrow l}$ interacting with $n-1$ ingoing fields is of the form \cite{Tezak2012}
\begin{eqnarray}
\left[G \right]_{k \rightarrow l} =\left(S_{k \rightarrow l}, L_{k \rightarrow l}, H_{k \rightarrow l}
\right) \label{eq:feedback} 
\end{eqnarray}
where
\begin{eqnarray}
S_{k \rightarrow l}&=&S_{\bar{k}, \bar{l}} - S_{\bar{k},l} (1- S_{jk})^{-1} S_{k, \bar{l}}, \nonumber\\
L_{k \rightarrow l}&=&L_{\bar{k}} - S_{\bar{k},l} (1- S_{jk})^{-1} L_k, \nonumber\\
H_{k \rightarrow l}&=&H+{\rm Im} \left( \left(\sum_{j=1}^{n}L_j^*S_{jl}\right)  (1- S_{jk})^{-1} L_k \right), \nonumber
\end{eqnarray}
$S_{\bar{k}, \bar{l}} $ is the matrix $S$ with the $k$th row and $l$th column removed, $S_{\bar{k},l}$ is the $l$th column of $S$ with the element $S_{kl}$ removed, $ S_{k, \bar{l}}$ is the $k$th row of $S$ with the element $S_{kl}$ removed, and $L_{\bar{k}} $ is the vector $L$ with the $k$th element $L_k$ removed.

%%%%%%%%%%%%%%%%%%%%%%%%%%%%%%%%%%%%%%%%%%%%%%%%%%%%%%%%%%%%%%%%%%%%%%%%%%%%%%%%%%%%%%%%%%%%%%%%%%%%%%%%%%%%%%%%%%%%%%%%%%%%%%%%%%%%%%%%%%%%%%%%%%%%%%%%%
%%%%%%%%%%%%%%%%%%%%%%%%%%%%%%%%%%%%%%%%%%%%%%%%%%%%%%%%%%%%%%%%%%%%%%%%%%%%%%%%%%%%%%%%%%%%%%%%%%%%%%%%%%%%%%%%%%%%%%%%%%%%%%%%%%%%%%%%%%%%%%%%%%%%%%%%%
\section{\label{sec:full_model} SLH models of elementary quantum components and cavity nonlinear optics}
In this section, we first give SLH parameters of basic quantum elements employed by the cavity nonlinear optics of interest. After that, we present SLH triples and master equations for a Kerr-cavity-based AND gate, NOT gate and NAND latch.
\begin{figure}[htbp]
	\begin{center}
		\includegraphics[scale=0.56]{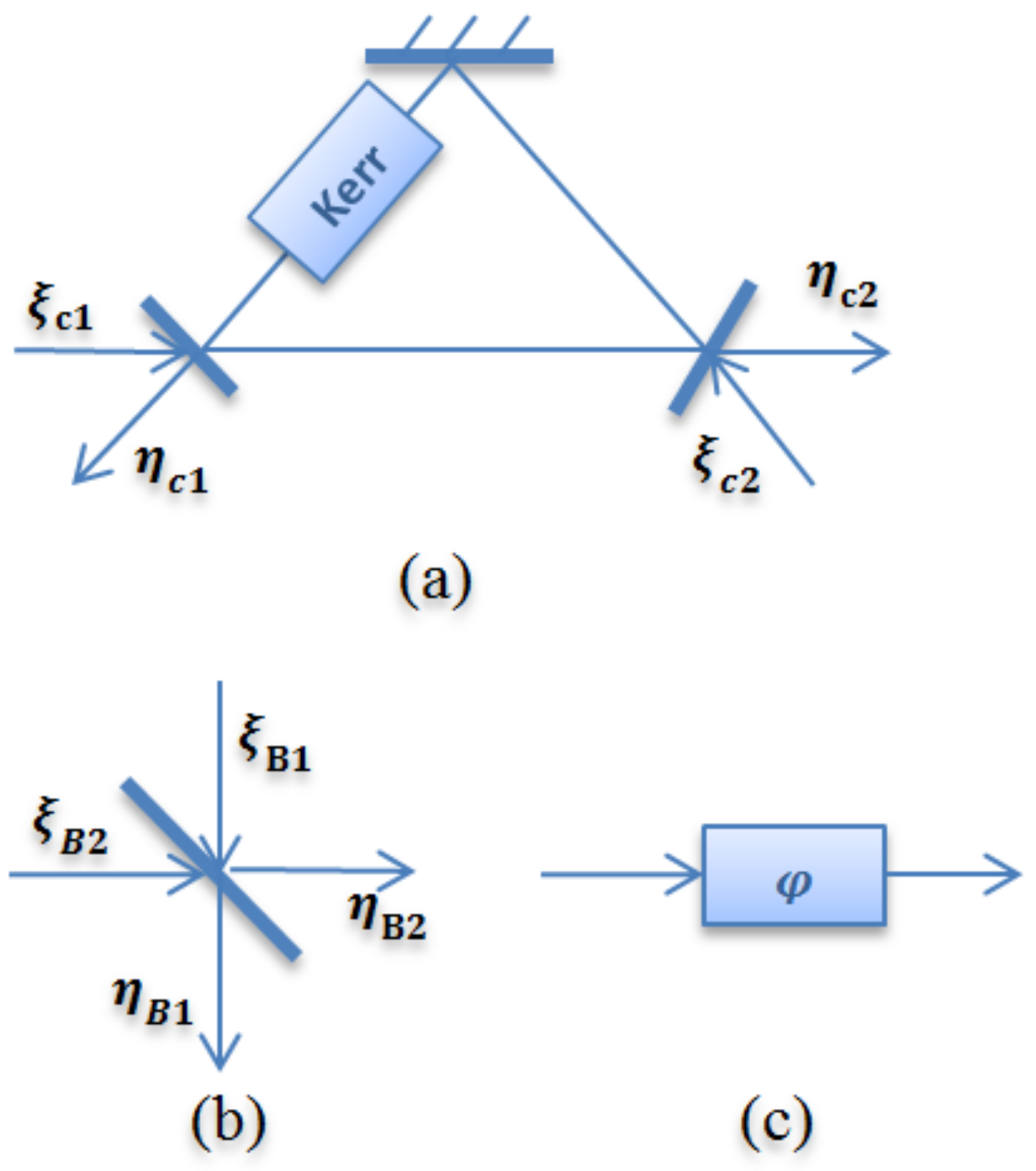}% Here is how to import EPS art
		\caption{\label{fig:optics} (a) A Kerr nonlinear optical ring cavity; (b) a beamsplitter; (c) a phase shifter.}
	\end{center}
\end{figure}

\subsection{\label{sec:elementary} Elementary quantum optical components}
\subsubsection{Kerr nonlinear optical ring cavities with vacuum ingoing fields}
A Kerr nonlinear optical ring cavity with two input/output channels is a system composed of a Kerr medium, two partially transmitting mirrors with damping rate $\kappa$ and a fully reflecting mirror, as shown in figure~\ref{fig:optics}(a). The Kerr medium has a second-order intensity dependent refractive index that causes nonlinearity in the relation between ingoing and outgoing optical fields \cite{bBachor2009}. 
The SLH model for the Kerr nonlinear cavity with two vacuum ingoing fields is
\begin{eqnarray}
K_a=\left(
I_2, 
\left( \begin{array}{c} \sqrt{\kappa}a \\ \sqrt{\kappa}a \end{array}\right),  H_0^a
\right), \label{eq:kerr}
\end{eqnarray}
where 
\begin{eqnarray}
H_0^a=\Delta a^*a+ \chi a^*a^*aa,
\end{eqnarray}
$\Delta$ is the detuning of the cavity resonance and $\chi$ is the Kerr nonlinear coefficient \cite{Mabuchi2011}.

For the sake of simplifying the derivation, we decompose $K_a$ into two concatenated subsystems $K_1^a$ and $K_1^b$ such that $K_a=K_1^a  \boxplus K_2^a $, where $K_1^a =\left(1, \sqrt{\kappa}a, 0 \right)$ and $K_2^a= \left(1, \sqrt{\kappa}a, \Delta a^*a+ \chi a^*a^*aa \right) $. For examples, see the derivations for SLH coefficients of the NOT gate and NAND latch in Section~\ref{sec:gates_app}.

\subsubsection{Beam splitters}
As figure~\ref{fig:optics}(b) shows, a beam splitter is a linear passive transformation (system energy is conserved) from two incoming fields to two outgoing fields. For a beam splitter with mixing angle $\theta$, its SLH model is
\begin{eqnarray}
B_\theta =\left(
\left( \begin{array}{cc} \cos(\theta) & -\sin(\theta) \\ \sin (\theta) & \cos(\theta) \end{array}\right),
O, 
0
\right). \label{eq:bs}
\end{eqnarray}

\subsubsection{ Phase shifters}
A phase shifter shown in figure~\ref{fig:optics}(c) is a linear quantum component that produces an angular shift $\varphi$ between its ingoing and outgoing fields. Its SLH triple is
\begin{eqnarray}
\Phi_\varphi =\left(
e^{\imath \varphi},
0, 
0
\right). \label{eq:ps}
\end{eqnarray}

\subsubsection{Coherent displacements}
A coherent ingoing field with amplitude function $\alpha$ can be generated as the output of the  SLH model 
\begin{eqnarray}
D_\alpha =\left(
1,
\alpha, 
0
\right). \label{eq:coheret}
\end{eqnarray}

\subsubsection{Channel permuting systems}
The series product requires one-to-one corresponding connections between ingoing channels of subsystem $G_2$ and output ports of $G_1$. That is, the $j$th outgoing field of $G_1$　 is fed into the $j$th input port of $G_2$. If connections between the subsystems do not follow such a sequence, a channel permuting system is required to reorder the outgoing fields of $G_1$ \cite{Tezak2012} (the ingoing fields of $G_2$). The permuting system is of the form
\begin{eqnarray}
P_{\sigma} =\left(
P,
O, 
0
\right), \label{eq:permuting}
\end{eqnarray}
where $\sigma$ is a vector specifies the permutation sequence and $P$ is the corresponding permutation matrix whose element is $P_{jk}=\delta_{j\sigma(k)}$.
For example, to change the outgoing fields sequence of a system $G$ with three outputs from $\eta_1, \eta_2, \eta_3$ to $\eta_1, \eta_3, \eta_2$, we employ a permuting system with $\sigma=[1,3,2]$ and 
\begin{eqnarray}
P = \left( \begin{array}{ccc} 
\delta_{11} & \delta_{13} & \delta_{12}\\
\delta_{21} & \delta_{23} & \delta_{22}\\
\delta_{31} & \delta_{33} & \delta_{32}
\end{array}\right)
=\left( \begin{array}{ccc} 
1 & 0 & 0\\
0 & 0 & 1\\
0 & 1 & 0
\end{array}\right). \nonumber
\end{eqnarray}

\subsubsection{Identity systems}
An identity system \cite{Tezak2012} with $n$ inputs/outputs is defined as
\begin{eqnarray}
\mathbf{1}_{n} =\left(
I_n,
O, 
0
\right). \label{eq:IDENTITY}
\end{eqnarray}
It passes its input straight through as its output.  

\subsection{\label{sec:gates_app} Cavity nonlinear optics}
In this section, we investigate SLH  models of Kerr nonlinear cavity optics, such as a Kerr nonlinear cavity with a coherent incoming field, an AND gate, a NOT gate and a NAND latch, as originally proposed in \cite{Mabuchi2011}. All the parameter values of the systems are the same as the ones in \cite{Mabuchi2011}.

\subsubsection{Kerr nonlinear cavities with coherent inputs}
The SLH model of a Kerr nonlinear ring cavity with a coherent input with amplitude $\epsilon$ fed into the first input channel and a vacuum field going through the second ingoing port is
\begin{eqnarray}
G_{K}&=&K_a \lhd \left(D_{\epsilon} \boxplus \mathbf{1}_1 \right)
=(S_K, L_K, H_K), \nonumber
\end{eqnarray}
where
\begin{eqnarray}
S_K&=&I_2, \nonumber\\
L_K&=&\left[\begin{array}{c}
\sqrt{\kappa}a+\epsilon\\
\sqrt{\kappa}a
\end{array}\right], \nonumber\\
H_K&=&H_0^a+\imath \frac{\sqrt{\kappa}}{2} \epsilon \left( a-a^* \right).
\end{eqnarray}

\subsubsection{AND gates}
\begin{figure}[htbp]
	\begin{center}
		\includegraphics[scale=0.5]{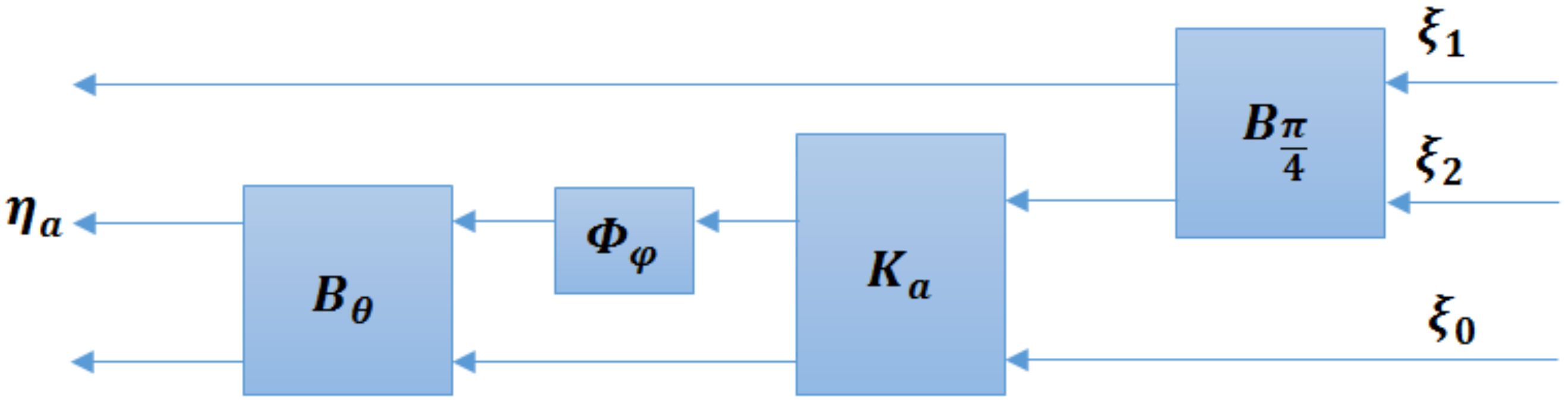}% Here is how to import EPS art
		\caption{\label{fig:AND} An AND gate setup.}
	\end{center}
\end{figure}
An AND gate setup is shown in figure~\ref{fig:AND}, with $\varphi=1.572$ and $\theta=1.073$. The gate has three incoming fields, among which $\xi_1$ and $\xi_2$ are coherent fields whose amplitude can be either $0$ (in this case the field is vacuum) or $\alpha=22.6274$, and $\xi_0$ is a vacuum field. 
The SLH coefficients for an AND gate are
\begin{eqnarray}
G_{A}&=&\left( \mathbf{1}_1 \boxplus \left(B_\theta \lhd  \left(\Phi_\varphi \boxplus \mathbf{1}_1 \right) \lhd K_a\right)\right) \lhd  \left( B_{\frac{\pi}{4}} \boxplus  \mathbf{1}_1 \right) \lhd \left(D_{\xi_1} \boxplus D_{\xi_2} \boxplus \mathbf{1}_1 \right)\nonumber\\
&=&(S_A, L_A, H_A), \nonumber
\end{eqnarray}
where
\begin{eqnarray}
S_A&=&\left[\begin{array}{ccc}
\frac{1}{\sqrt{2}} & -\frac{1}{\sqrt{2}} & 0 \\
\frac{1}{\sqrt{2}} \cos\theta e^{\imath \varphi} &\frac{1}{\sqrt{2}} \cos\theta e^{\imath \varphi}& -\sin \theta\\
\frac{1}{\sqrt{2}} \sin\theta e^{\imath \varphi} &\frac{1}{\sqrt{2}} \sin\theta e^{\imath \varphi}& \cos \theta\\
\end{array}\right], \nonumber\\
L_A&=&\left[\begin{array}{c}
\frac{1}{\sqrt{2}}\left(\xi_1-\xi_2\right)\\
L_{A1}\\
L_{A2}
\end{array}\right], \nonumber\\
L_{A1}&=&\cos\theta e^{\imath \varphi}\sqrt{\kappa}a- \sin\theta\sqrt{\kappa}a+\frac{1}{\sqrt{2}}\cos\theta e^{\imath \varphi}\left(\xi_1+\xi_2\right),\nonumber\\
L_{A2}&=&\sin\theta e^{\imath \varphi}\sqrt{\kappa}a+ \cos\theta\sqrt{\kappa}a+\frac{1}{\sqrt{2}}\sin\theta e^{\imath \varphi}\left(\xi_1+\xi_2\right),\nonumber\\
H_A&=&H_0^a+\imath \frac{\sqrt{\kappa}}{2\sqrt{2}} \left(\xi_1+\xi_2\right) \left( a-a^* \right).
\end{eqnarray}
The master equation of the AND gate is
\begin{eqnarray}
\frac{d}{dt}\rho_{a} =-\imath \left[H_{a}, \rho_{a} \right] + 2\left(L_{a} \rho L_{a}^* - \frac{1}{2} L_{a}^*L_{a}\rho_{a} - \frac{1}{2} \rho_{a} L_{a}^*L_{a} \right), \nonumber
\end{eqnarray}
where
\begin{eqnarray}
H_a &=&H_0+\imath \sqrt{\frac{\kappa}{2}} \left(\xi_1+\xi_2\right) \left( a-a^* \right), \nonumber\\
L_a &=& \sqrt{\kappa} a.
\end{eqnarray}
The expectation value of the output field of interest is 
\begin{eqnarray}
\left\langle\eta_{a}\right\rangle&=&\cos\theta e^{\imath \varphi}\sqrt{\kappa}a- \sin\theta\sqrt{\kappa}a 
+\frac{1}{\sqrt{2}}\cos\theta e^{\imath \varphi}\left(\xi_1+\xi_2\right).\nonumber
\end{eqnarray} 

\subsubsection{NOT gates}
\begin{figure}[htbp]
	\begin{center}
		\includegraphics[scale=0.38]{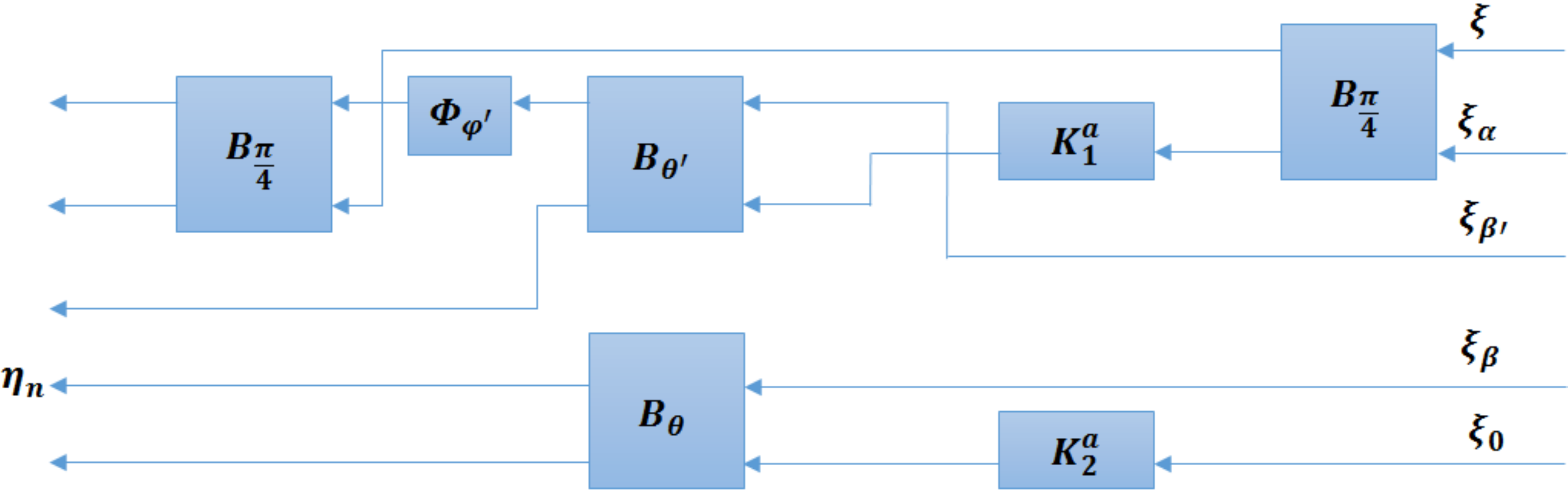}% Here is how to import EPS art
		\caption{\label{fig:NOT} A NOT gate setup.}
	\end{center}
\end{figure}
A NOT gate as shown in figure~\ref{fig:NOT} has five incoming fields, among which three coherent fields $\xi_\alpha$, $\xi_\beta$ and $\xi_{\beta'}$ have fixed amplitudes $\alpha$, $\beta$ and $\beta'$, respectively, $\xi_0$ is a vacuum field, and $\xi$ is an adjustable input. The system parameters are set as follows, $\theta=0.891$, $\theta'=1.071$, $\varphi'=2.03$, $\beta=-34.289-11.909\imath$, $\beta'=7.833-17.656 \imath$ and $\alpha=22.6274$.
The SLH model for a NOT gate is
\begin{eqnarray}
&&G_{N}= \left(\left(\left(B_{\frac{\pi}{4}} \boxplus \mathbf{1}_1 \right) \lhd P_{[2,1,3]} \lhd \left(\mathbf{1}_1  \boxplus \left( \left(\Phi_{\varphi'}\right) \lhd B_{\theta'} \lhd  P_{[2,1]}\right) \right)  \right. \right. \nonumber\\
&&\quad \quad \left. \left. \lhd \left( \left(\left(\mathbf{1}_1 \boxplus K_1^a\right) \lhd B_{\frac{\pi}{4}} \right) \boxplus \mathbf{1}_1  \right)  \right) \boxplus \left(B_{\theta}  \lhd \left(\mathbf{1}_1  \boxplus K_2^a \right)\right) \right)\nonumber\\
&&\quad \quad \lhd \left(D_{\xi}  \boxplus D_{\alpha} \boxplus D_{\beta'} \boxplus D_{\beta} \boxplus  \mathbf{1}_1 \right)\nonumber\\
&&\quad \quad =(S_N, L_N, H_N),\nonumber
\end{eqnarray}
\normalsize
where
\begin{eqnarray}
S_N&=&\left[\begin{array}{cc}
S_{N1}&O\\
O& S_{N2}
\end{array}\right], \nonumber\\
S_{N1}&=&\left[\begin{array}{ccc}
\frac{-1-e^{\imath\varphi'}\sin \theta'}{2} & \frac{1-e^{\imath\varphi'}\sin \theta'}{2} & \frac{e^{\imath \varphi'}\cos \theta'}{\sqrt{2}}\\
\frac{1-e^{\imath\varphi'}\sin \theta'}{2}  & \frac{-1-e^{\imath\varphi'}\sin \theta'}{2} & \frac{e^{\imath \varphi'}\cos \theta'}{\sqrt{2}}\\
\frac{\cos \theta'}{\sqrt{2}} & \frac{\cos \theta'}{\sqrt{2}} & \sin \theta'
\end{array}\right], \nonumber\\
S_{N2}&=&\left[\begin{array}{ccc}
\cos \theta & -\sin \theta \\
\sin \theta & \cos \theta
\end{array}\right], \nonumber\\
L_N&=&\left[\begin{array}{c}
L_{N1}\\
L_{N2}\\
\sqrt{\kappa}a \cos \theta' + \frac{\cos \theta'}{\sqrt{2}}\left(\xi +\alpha\right)+\sin \theta' \beta'\\
-\sqrt{\kappa}a\sin \theta +\beta \cos \theta\\
\sqrt{\kappa}a\cos \theta +\beta \sin \theta\\
\end{array}\right],\nonumber\\
H_N&=& H_0^a+\imath \frac{1}{2} \sqrt{\frac{\kappa}{2} }\left( a-a^* \right)\left(\xi + \alpha \right), \nonumber\\
L_{N1}&=&-\frac{\sqrt{\kappa}a e^{\imath\varphi'}\sin \theta'}{\sqrt{2}} +\frac{-1-e^{\imath\varphi'}\sin \theta'}{2} \xi+\frac{1-e^{\imath\varphi'}\sin \theta'}{2}   \alpha  +\frac{ e^{\imath\varphi'}\cos \theta'}{\sqrt{2}} \beta',\nonumber\\
L_{N2}&=&-\frac{\sqrt{\kappa}a e^{\imath\varphi'}\sin \theta'}{\sqrt{2}} +\frac{1-e^{\imath\varphi'}\sin \theta'}{2}  \xi+\frac{-1-e^{\imath\varphi'}\sin \theta'}{2}   \alpha 
+\frac{ e^{\imath\varphi'}\cos \theta'}{\sqrt{2}} \beta'.
\end{eqnarray}
\normalsize

The master equation of the NOT gate is
\begin{eqnarray}
\frac{d}{dt}\rho_{n} =-\imath \left[H_{n}, \rho_{n} \right] + 2\left(L_{n} \rho_{n} L_{n}^* - \frac{1}{2} L_{n}^*L_{n}\rho_{n} - \frac{1}{2} \rho_{n} L_{n}^*L_{n} \right), \nonumber
\end{eqnarray}
\normalsize
where
\begin{eqnarray}
H_n &=&\Delta a^*a+\chi a^*a^*aa+\imath \sqrt{\frac{\kappa}{2}} \left(\alpha+\xi \right) \left( a-a^* \right), \nonumber\\
L_n &=& \sqrt{\kappa} a.
\end{eqnarray}
The expectation value of the output field of interest is 
\begin{eqnarray}
\left\langle\eta_{n}\right\rangle=\beta \cos\theta  -\sqrt{\kappa}a \sin\theta. \nonumber
\end{eqnarray}

\subsubsection{NAND latches}
\begin{figure}[htbp]
	\begin{center}
		\includegraphics[scale=0.55]{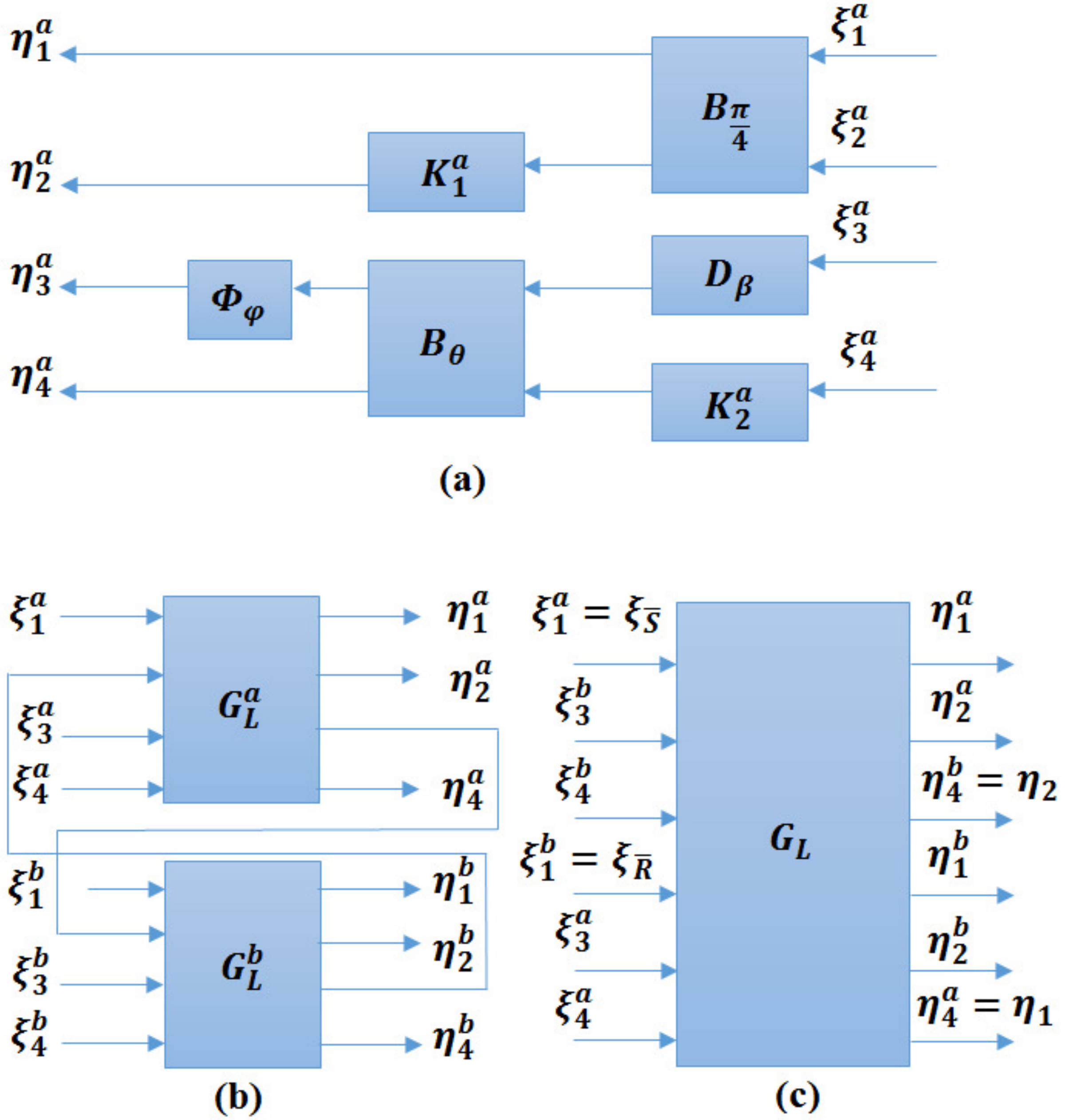}% Here is how to import EPS art
		\caption{\label{fig:NAND} A NAND gate setup. (a) Subsystem $G_{L}^a$. (b)  $\left[\left[ G_{L}^a \boxplus G_{L}^b\right]_{3 \rightarrow 6}\right]_{6 \rightarrow 2} $, feedback connections between subsystems $G_{L}^a$ and $G_{L}^b$. (c) A NAND gate $G_{L}$}.
	\end{center}
\end{figure}
A NAND latch as shown in figure~\ref{fig:NAND}(c)  is a dual-Kerr-cavity system. It has two identical Kerr-cavity subsystems $G_{L}^a$ and $G_{L}^b$ as shown in figure~\ref{fig:NAND}(a)  with parameters $\beta=-34.289-11.909 \imath$, $\theta=0.891$ and $\varphi =2.546$, which are connected in a coherent feedback loop, see figure~\ref{fig:NAND}(b).  The NAND latch has $6$ incoming fields, where two inputs $\xi_{\bar S}$ and $\xi_{\bar R}$ are coherent fields with adjustable amplitudes $\bar S$ and $\bar R$ respectively that can be either $0$ or $\alpha$. The other fields $\xi_3^b$, $\xi_4^b$, $\xi_3^a$ and $\xi_4^a$ shown in figure~\ref{fig:NAND}(c) are vacuum fields. 
The SLH model for the NAND latch is
\begin{eqnarray}
G_{L}&=& P_{[1,2,6, 4,5,3 ]} \lhd\left[\left[ G_{L}^a \boxplus G_{L}^b\right]_{3 \rightarrow 6}\right]_{6 \rightarrow 2} \lhd  P_{[1,5,6, 4,2,3 ]}   \lhd \left( \bar S \boxplus \mathbf{1}_2 \boxplus  \bar R\boxplus \mathbf{1}_2  \right) \nonumber\\
&=&(S_L, L_L, H_L),\nonumber
\end{eqnarray}
where
\begin{eqnarray}
&& G_{L}^a=\left( \left( \mathbf{1}_1 \boxplus K_{1}^a \right) \lhd B_{\frac{\pi}{4}}\right) \boxplus \left(\left(\Phi_{\varphi} \boxplus \mathbf{1}_1 \right) \lhd B_\theta \lhd \left(D_{\beta}   \boxplus K_2^a\right)\right), \nonumber\\
&& G_{L}^b=\left( \left( \mathbf{1}_1 \boxplus K_{1}^b \right) \lhd B_{\frac{\pi}{4}}\right) \boxplus \left(\left(\Phi_{\varphi} \boxplus \mathbf{1}_1 \right) \lhd B_\theta \lhd \left(D_{\beta}\boxplus K_2^b\right)\right), \nonumber\\
&& S_L= I_2 \otimes S_1 \nonumber\\
&& S_1 =\left[\begin{array}{ccc}
\frac{1}{\sqrt{2}} & -\frac{1}{\sqrt{2}} \cos\theta e^{\imath \varphi} &  \frac{1}{\sqrt{2}} \sin\theta e^{\imath \varphi} \\
\frac{1}{\sqrt{2}} &  \frac{1}{\sqrt{2}} \cos\theta e^{\imath \varphi} & -\frac{1}{\sqrt{2}} \sin\theta e^{\imath \varphi}\\
0                  &  \sin\theta                                    & \cos \theta
\end{array}\right], \nonumber\\
&& L_L=\left[\begin{array}{c}
\sqrt{\frac{\kappa}{2}}\sin\theta e^{\imath \varphi}b- \frac{\beta}{\sqrt{2}}\cos \theta e^{\imath \varphi} +\frac{\bar{S}}{\sqrt{2}}\\
\sqrt{\kappa}a-\sqrt{\frac{\kappa}{2}}\sin\theta e^{\imath \varphi}b+ \frac{\beta}{\sqrt{2}}\cos \theta e^{\imath \varphi} +\frac{\bar{S}}{\sqrt{2}}\\
\sqrt{\kappa}\cos\theta b+\beta\sin \theta\\
\sqrt{\frac{\kappa}{2}}\sin\theta e^{\imath \varphi}a- \frac{\beta}{\sqrt{2}}\cos \theta e^{\imath \varphi} +\frac{\bar{R}}{\sqrt{2}}\\
\sqrt{\kappa}b-\sqrt{\frac{\kappa}{2}}\sin\theta e^{\imath \varphi}a+ \frac{\beta}{\sqrt{2}}\cos \theta e^{\imath \varphi} +\frac{\bar{R}}{\sqrt{2}}\\
\sqrt{\kappa}\cos\theta a+\beta\sin \theta
\end{array}\right], \nonumber\\
&&H_L=H_{0}^a+H_{0}^b- \frac{\kappa}{\sqrt{2}}\sin \theta \sin \varphi \left(ab^*+a^*b\right)  +\imath \frac{\sqrt{\kappa}}{2\sqrt{2}} \left\lbrace\left(\bar{S}^*+\beta^*\cos \theta e^{-\imath \varphi}\right)a \right. \nonumber\\
&& \quad \left. - \left(\bar{S}+\beta\cos \theta e^{\imath \varphi}\right)a^* +\left(\bar{R}^*+\beta^*\cos \theta e^{-\imath \varphi}\right)b - \left(\bar{R}+\beta\cos \theta e^{\imath \varphi}\right)b^*  \right\rbrace. 
\end{eqnarray}

The master equation of the NAND latch is
\begin{eqnarray}
\frac{d}{dt}\rho_l =-\imath \left[H_l, \rho_l \right] + \sum_{j=1}^{4}\left(L_{j} \rho_l L_{j}^* - \frac{1}{2} L_{j}^*L_{j}\rho_l  - \frac{1}{2} \rho_l L_j^*L_j \right), \nonumber
\end{eqnarray}
where
\begin{eqnarray}
H_l &=&\Delta \left(a^*a+b^*b\right)+\chi \left(a^*a^*aa+b^*b^*bb\right)  - \frac{\kappa}{\sqrt{2}}\sin \theta \sin \varphi \left(ab^*+a^*b\right) \nonumber\\
&& \quad
+\imath \sqrt{\frac{\kappa}{2}}  \left\lbrace\left(\bar{S}^*+\beta^*\cos \theta e^{-\imath \varphi}\right)a - \left(\bar{S}+\beta\cos \theta e^{\imath \varphi}\right)a^* \right. \nonumber\\
&& \quad+\left. \left(\bar{R}^*+\beta^*\cos \theta e^{-\imath \varphi}\right)b - \left(\bar{R}+\beta\cos \theta e^{\imath \varphi}\right)b^*  \right\rbrace, \nonumber \\
L_1 &=&\sqrt{\frac{\kappa}{2}\left(1+\cos^2\theta \right)} a, \nonumber\\
L_2&=&\sqrt{\frac{\kappa}{2}}\sin \theta e^{\imath\varphi}a-\sqrt{\kappa}b, \nonumber
\end{eqnarray}
\begin{eqnarray}
L_3 &=& \sqrt{\frac{\kappa}{2}\left(1+\cos^2\theta \right)} b, \nonumber\\
L_4 &=& \sqrt{\frac{\kappa}{2}}\sin \theta e^{\imath\varphi}b-\sqrt{\kappa}a.
\end{eqnarray}
%
%The expectation value of the output field of interest is 
%\begin{eqnarray}
%\left\langle\eta_{l}\right\rangle=\cos\theta'\left\langle\eta_{2}\right\rangle-\sin\theta'\left\langle\eta_{1}\right\rangle e^{\imath \varphi'},
%\end{eqnarray}
%where $\theta'=0.566$, $\varphi'=0.158$, 
%\begin{eqnarray}
%\left\langle\eta_1\right\rangle&=&\sqrt{\kappa}\cos\theta a+\beta\sin \theta, \nonumber\\
%\left\langle\eta_2\right\rangle&=&\sqrt{\kappa}\cos\theta b+\beta\sin \theta. \nonumber
%%\end{eqnarray}

%%%%%%%%%%%%%%%%%%%%%%%%%%%%%%%%%%%%%%%%%%%%%%%%%%%%%%%%%%%%%%%%%%%%%%%%%%%%%%%%%%%%%%%%%%%%%%%%%%%%%%%%%%%%%%%%%%%%%%%%%%%%%%%%%%%%%%%%%%%%%%%%%%%%%%%%%
%%%%%%%%%%%%%%%%%%%%%%%%%%%%%%%%%%%%%%%%%%%%%%%%%%%%%%%%%%%%%%%%%%%%%%%%%%%%%%%%%%%%%%%%%%%%%%%%%%%%%%%%%%%%%%%%%%%%%%%%%%%%%%%%%%%%%%%%%%%%%%%%%%%%%%%%%
\section{\label{sec:reduced_model} Reduced models of cavity nonlinear optics}
In this section we describe the reduced model for the Kerr nonlinear cavity following the strategy outlined in the main text, and then obtain reduced models for the AND and NOT gates, and the NAND latch, by connecting the reduced cavity model with the remaining optical components in the network realizing the gates and latch. 
\subsection{\label{sec:kerr_coherent_reduce} Reduced model for Kerr nonlinear cavities}
The Hilbert space for the reduced Kerr-cavity model is the span of the last $d$ columns of the matrix $T$ as described in the text. The SLH model of a Kerr nonlinear ring cavity with a coherent input $\epsilon$ fed into the first input channel and a vacuum field going through the second ingoing port is
\begin{eqnarray}
G_{K}^{(r)}=(S_K^{(r)}, L_K^{(r)}, H_K^{(r)}), \nonumber
\end{eqnarray}
where
\begin{eqnarray}
S_K^{(r)}&=&S_K, \nonumber\\
L_K^{(r)}&=&\left[\begin{array}{c}
\sqrt{\kappa}a^{(r)}+\epsilon\\
\sqrt{\kappa}a^{(r)}
\end{array}\right], \nonumber\\
H_K^{(r)}&=&PT^* \left(H_0^a+\imath \frac{\sqrt{\kappa}}{2} \epsilon \left( a-a^* \right) \right)TP^* \nonumber\\
&=&PT^* H_0^aTP^* +\imath \frac{\sqrt{\kappa}}{2} \epsilon \left( PT^*aTP^*-PT^*a^*TP^* \right) \nonumber\\
&=&PT^* H_0^aTP^* +\imath \frac{\sqrt{\kappa}}{2} \epsilon \left( PT^*aTP^*-\left(PT^*aTP^*\right)^* \right) \nonumber\\
&=&{H_0^a}^{(r)}+\imath \frac{\sqrt{\kappa}}{2} \epsilon \left( a^{(r)}-{a^{(r)}}^* \right) \nonumber
\end{eqnarray}
with
\begin{eqnarray}
a^{(r)}=PT^* a TP^*,
~~ {H_{0}^a}^{(r)}=PT^* H_0^aTP^*. \label{eq:reduced_kerr_coherent}
\end{eqnarray}

In particular, when both cavity ingoing field are vacuum ($\epsilon=0$), following (\ref{eq:reduced_kerr_coherent}),  the reduced SLH model for a Kerr nonlinear cavity with vacuum ingoing fields shown as (\ref{eq:kerr}) is 
\begin{eqnarray}
K_a^{(r)} =\left(
I_2, 
\left( \begin{array}{c} \sqrt{\kappa}a^{(r)} \\ \sqrt{\kappa}a^{(r)} \end{array}\right),  {H_0^a}^{(r)}
\right). \label{eq:reduced_kerr}
\end{eqnarray}
Furthermore, we can write
\begin{eqnarray}
K_a^{(r)}={K_1^a}^{(r)} \boxplus{K_2^a}^{(r)} , \nonumber
\end{eqnarray}
with
\begin{eqnarray}
{K_1^a}^{(r)}&=&(1, \sqrt{\kappa}a^{(r)},0), \nonumber\\
{K_2^a}^{(r)} &=&(1, \sqrt{\kappa}a^{(r)}, {H_0^a}^{(r)}). 
\end{eqnarray}

\subsection{\label{sec:and_reduce} Reduced model of AND gates}
The SLH model for a reduced AND gate is
\begin{eqnarray}
G_{A}^{(r)}&=&\left( \mathbf{1}_1 \boxplus \left(B_\theta \lhd  \left(\Phi_\varphi \boxplus \mathbf{1}_1 \right) \lhd K_a^{(r)}\right)\right) \lhd  \left( B_{\frac{\pi}{4}} \boxplus  \mathbf{1}_1 \right) \lhd \left(D_{\xi_1} \boxplus D_{\xi_2} \boxplus \mathbf{1}_1 \right)\nonumber\\
&=&(S_A^{(r)}, L_A^{(r)}, H_A^{(r)}), \nonumber
\end{eqnarray}
where
\begin{eqnarray}
S_A^{(r)}&=&S_A, \nonumber\\
L_A^{(r)} &=&\left[\begin{array}{c}
\frac{1}{\sqrt{2}}\left(\xi_1-\xi_2\right)\\
L_{A1}^{(r)}\\
L_{A2}^{(r)}
\end{array}\right], \nonumber\\
L_{A1}^{(r)}&=&(\cos\theta e^{\imath \varphi}- \sin\theta)\sqrt{\kappa}a^{(r)}+\frac{1}{\sqrt{2}}\cos\theta e^{\imath \varphi}\left(\xi_1+\xi_2\right),\nonumber\\
L_{A2}^{(r)}&=& (\sin\theta e^{\imath \varphi}+ \cos\theta)\sqrt{\kappa}a^{(r)}+\frac{1}{\sqrt{2}}\sin\theta e^{\imath \varphi}\left(\xi_1+\xi_2\right), \nonumber\\
H_A^{(r)} &=&{H_{0}^a}^{(r)} +\imath \frac{\sqrt{\kappa}}{2\sqrt{2}} \left(\xi_1+\xi_2\right) \left( a^{(r)} -{a^{(r)} }^* \right).
\end{eqnarray}
The expectation value of the output field of interest is 
\begin{eqnarray}
\left\langle\eta_{a}^{(r)}\right\rangle&=&\cos\theta e^{\imath \varphi}\sqrt{\kappa}a^{(r)}- \sin\theta\sqrt{\kappa}a^{(r)} 
+\frac{1}{\sqrt{2}}\cos\theta e^{\imath \varphi}\left(\xi_1+\xi_2\right). \nonumber
\end{eqnarray}

\subsection{\label{sec:not_reduce} Reduced model of NOT gates}
The SLH model for a reduced NOT gate system is
\begin{eqnarray}
G_{N}^{(r)} &=& \left(\left(\left(B_{\frac{\pi}{4}} \boxplus \mathbf{1}_1 \right) \lhd P_{[2,1,3]} \lhd \left(\mathbf{1}_1  \boxplus \left( \left(\Phi_{\varphi'}\right) \lhd B_{\theta'} \lhd  P_{[2,1]}\right) \right) \right. \right. \nonumber\\
&&\left. \lhd \left( \left(\left(\mathbf{1}_1 \boxplus {K_1^a}^{(r)} \right) \lhd B_{\frac{\pi}{4}} \right) \boxplus \mathbf{1}_1  \right)  \right)  \left. \boxplus\left(B_{\theta}  \lhd \left(\mathbf{1}_1  \boxplus {K_2^a}^{(r)}  \right)\right) \right)\nonumber\\
&&\lhd \left(D_{\xi}  \boxplus D_{\alpha} \boxplus D_{\beta'} \boxplus D_{\beta} \boxplus  \mathbf{1}_1 \right)\nonumber\\
&=&(S_N^{(r)}, L_N^{(r)}, H_N^{(r)}), \nonumber
\end{eqnarray}
where
\begin{eqnarray}
S_N^{(r)}&=&S_N, \nonumber\\
L_N^{(r)}&=&\left[\begin{array}{c}
L_{N1}^{(r)}\\
L_{N2}^{(r)}\\
\sqrt{\kappa}a^{(r)} \cos \theta' + \frac{\cos \theta'}{\sqrt{2}}\left(\xi +\alpha\right)+\sin \theta' \beta'\\
-\sqrt{\kappa}a^{(r)}\sin \theta +\beta \cos \theta\\
\sqrt{\kappa}a^{(r)}\cos \theta +\beta \sin \theta\\
\end{array}\right],\nonumber\\
H_N^{(r)}&=& {H_{0}^a}^{(r)}+\imath \frac{1}{2} \sqrt{\frac{\kappa}{2} }\left( a^{(r)}-{a^{(r)}}^* \right)\left(\xi+ \alpha \right), \nonumber\\
L_{N1}^{(r)}&=&-\frac{\sqrt{\kappa}a^{(r)} e^{\imath\varphi'}\sin \theta'}{\sqrt{2}} +\frac{-1-e^{\imath\varphi'}\sin \theta'}{2}  \xi+\frac{1-e^{\imath\varphi'}\sin \theta'}{2}   \alpha+\frac{ e^{\imath\varphi'}\cos \theta'}{\sqrt{2}} \beta' \nonumber\\
L_{N2}^{(r)}&=&-\frac{\sqrt{\kappa}a^{(r)} e^{\imath\varphi'}\sin \theta'}{\sqrt{2}} +\frac{1-e^{\imath\varphi'}\sin \theta'}{2}  \xi+\frac{-1-e^{\imath\varphi'}\sin \theta'}{2}   \alpha+\frac{ e^{\imath\varphi'}\cos \theta'}{\sqrt{2}} \beta'.
\end{eqnarray}
\normalsize
The expectation value of the output field of interest is 
\begin{eqnarray}
\left\langle\eta_{n}^{(r)}\right\rangle=\beta \cos\theta  -\sqrt{\kappa}a^{(r)} \sin\theta. \nonumber
\end{eqnarray}

\subsection{\label{sec:nand_reduce} Reduced model of NAND gates}
The SLH coefficients for a reduced larch model are
\begin{eqnarray}
G_{L}^{(r)}&=&P_{[1,2,6, 4,5,3 ]} \lhd\left[\left[ {G_{L}^a}^{(r)} \boxplus {G_{L}^b}^{(r)}\right]_{3 \rightarrow 6}\right]_{6 \rightarrow 2}  \lhd  P_{[1,5,6, 4,2,3 ]}  \lhd \left( \bar S \boxplus \mathbf{1}_2 \boxplus  \bar R\boxplus \mathbf{1}_2  \right) \nonumber\\
&=&(S_L^{(r)} , L_L^{(r)} , H_L^{(r)} ),\nonumber
\end{eqnarray}
where
\begin{eqnarray}
&&{G_{L}^a}^{(r)}=\left( \left( \mathbf{1}_1 \boxplus {K_1^a}^{(r)}  \right) \lhd B_{\frac{\pi}{4}}\right) \boxplus \left(\left(\Phi_{\varphi} \boxplus \mathbf{1}_1 \right) \lhd B_\theta   \lhd \left(D_{\beta} \boxplus {K_2^a}^{(r)} \right)\right), \nonumber\\
&& {G_{L}^b}^{(r)}=\left( \left( \mathbf{1}_1 \boxplus {K_1^b}^{(r)} \right) \lhd B_{\frac{\pi}{4}}\right) \boxplus \left(\left(\Phi_{\varphi} \boxplus \mathbf{1}_1 \right) \lhd B_\theta  \lhd \left(D_{\beta} \boxplus {K_2^b}^{(r)} \right)\right), \nonumber\\
&& S_L^{(r)} = S_L, \nonumber\\
&& L_L^{(r)}=\left[\begin{array}{c}
\sqrt{\frac{\kappa}{2}}\sin\theta e^{\imath \varphi}b^{(r)}- \frac{\beta}{\sqrt{2}}\cos \theta e^{\imath \varphi} +\frac{\bar{S}}{\sqrt{2}}\\
\sqrt{\kappa}a^{(r)}-\sqrt{\frac{\kappa}{2}}\sin\theta e^{\imath \varphi}b^{(r)}+ \frac{\beta}{\sqrt{2}}\cos \theta e^{\imath \varphi} +\frac{\bar{S}}{\sqrt{2}}\\
\sqrt{\kappa}\cos\theta b^{(r)}+\beta\sin \theta\\
\sqrt{\frac{\kappa}{2}}\sin\theta e^{\imath \varphi}a^{(r)}- \frac{\beta}{\sqrt{2}}\cos \theta e^{\imath \varphi} +\frac{\bar{R}}{\sqrt{2}}\\
\sqrt{\kappa}b^{(r)}-\sqrt{\frac{\kappa}{2}}\sin\theta e^{\imath \varphi}a^{(r)}+ \frac{\beta}{\sqrt{2}}\cos \theta e^{\imath \varphi} +\frac{\bar{R}}{\sqrt{2}}\\
\sqrt{\kappa}\cos\theta a^{(r)}+\beta\sin \theta
\end{array}\right], \nonumber\\
&&H_L^{(r)}={H_{0}^a}^{(r)}+{H_{0}^b}^{(r)}- \frac{\kappa}{\sqrt{2}}\sin \theta \sin \varphi \left(a^{(r)}{b^{(r)}}^* +{a^{(r)}}^*b^{(r)}\right) \nonumber\\
&& \quad\quad+\imath \frac{\sqrt{\kappa}}{2\sqrt{2}} \left\lbrace\left(\bar{S}^*+\beta^*\cos \theta e^{-\imath \varphi}\right)a^{(r)} - \left(\bar{S}+\beta\cos \theta e^{\imath \varphi}\right){a^{(r)}}^*  \right. \nonumber\\
&&\quad\quad \left.+\left(\bar{R}^*+\beta^*\cos \theta e^{-\imath \varphi}\right)b^{(r)} - \left(\bar{R}+\beta\cos \theta e^{\imath \varphi}\right){b^{(r)}}^*  \right\rbrace.\nonumber
\end{eqnarray}
\normalsize

%%%%%%%%%%%%%%%%%%%%%%%%%%%%%%
%%%%%%%%%%%%%%%%%%%%%%%%%%%%%%
\bibliography{refs} 
\bibliographystyle{unsrt}
\end{document}